\documentclass[aps,pre,twocolumn,showpacs,superscriptaddress]{revtex4}

\usepackage{graphicx}% Include figure files
\usepackage{dcolumn}% Align table columns on decimal point
\usepackage{bm}% bold math
\usepackage{hyperref}% add hypertext capabilities
\usepackage{amsmath}
\usepackage[usenames]{color}
\begin{document}

\title{Range-limited Centrality Measures in Complex Networks }

\author{M\'aria Ercsey-Ravasz} \email{mercseyr@nd.edu}
\affiliation{Faculty of Physics, Babe\c{s}-Bolyai University, 
Str. Kogalniceanu Nr. 1, RO-400084 Cluj-Napoca, Romania}
\affiliation{Interdisciplinary Center for Network Science and 
Applications (iCeNSA), University of Notre Dame, Notre Dame, IN, 46556 USA}  
\author{Ryan Lichtenwalter}
\affiliation{Interdisciplinary Center for Network Science and 
Applications (iCeNSA), University of Notre Dame, Notre Dame, IN, 46556 USA}  
\affiliation{Department of Computer Science and Engineering, 
University of Notre Dame, Notre Dame, IN, 46556 USA}
\author{Nitesh V.  Chawla}
\affiliation{Interdisciplinary Center for Network Science and 
Applications (iCeNSA), University of Notre Dame, Notre Dame, IN, 46556 USA}  
\affiliation{Department of Computer Science and Engineering, 
University of Notre Dame, Notre Dame, IN, 46556 USA}
\author{Zolt\'an Toroczkai} \email{toro@nd.edu}
\affiliation{Interdisciplinary Center for Network Science and 
Applications (iCeNSA), University of Notre Dame, Notre Dame, IN, 46556 USA}  
\affiliation{Department of Computer Science and Engineering, 
University of Notre Dame, Notre Dame, IN, 46556 USA}
\affiliation{Department of Physics,
University of Notre Dame, Notre Dame, IN, 46556 USA}

\date{\today}

\begin{abstract}
Here we present a range-limited approach to centrality 
measures in both non-weighted and weighted directed complex networks. 
We introduce an efficient method that generates for every node and
every edge its betweenness centrality based on shortest paths of lengths
not longer than $\ell = 1,\ldots,L$ in case of non-weighted networks, and
for weighted networks the corresponding quantities based on 
minimum weight paths with path weights not larger than $w_{\ell}=\ell \Delta$, 
$\ell=1,2\ldots,L=R/\Delta$. These measures provide a systematic
description on the positioning importance of a node (edge) with respect to its network 
neighborhoods 1-step out, 2-steps out, etc. up to including the whole network. 
They are more informative than traditional centrality measures, as network transport 
typically happens on all length-scales, from transport to nearest neighbors to 
the farthest reaches of the network.  We show that
range-limited centralities obey universal scaling laws for large non-weighted networks. 
As the computation of traditional centrality measures is costly, this scaling behavior 
can be exploited to efficiently estimate centralities of nodes and edges for all ranges, 
including the traditional ones. The scaling behavior can also be exploited to show
that the ranking top-list of nodes (edges) based on their range-limited centralities 
quickly freezes as function of the range, and hence the diameter-range top-list 
can be efficiently predicted.  We also show how to estimate the typical largest 
node-to-node distance for a network of $N$ nodes, exploiting the afore-mentioned scaling behavior. 
These observations are illustrated on model networks and on a large social  network inferred from 
cell-phone trace logs ($\sim 5.5\times 10^6$ nodes and $\sim 2.7\times 10^7$ edges). Finally,
we apply these concepts to efficiently detect the vulnerability backbone
of a network (defined as the smallest percolating cluster of the highest betweenness 
nodes and edges) and illustrate the importance of 
weight-based centrality measures in weighted networks in detecting such backbones.
\end{abstract}


\maketitle

\section{Introduction}

Network research 
\cite{CohenHavlinBook,NewmanBook,BarratVespBook,BoccaEtAlRev06,BenFrauTor04} 
has experienced an explosive growth 
in the last two decades, as it has proven itself to be an informative and useful
methodology to study complex systems, ranging from social sciences 
through biology to communication infrastructures. Both the natural and man made 
world is abundant with networked structures that transport various entities, such as 
information, forces, energy, material goods, etc. As many of these
networks are the result of evolutionary processes, it is  important  to understand how the
graph structure of these systems determines their transport performance, structural stability and 
behavior as a whole.  A rather useful concept in addressing such questions is the notion
of centrality, which describes the positioning ``importance'' of a structure of interest such 
as a node, edge or subgraph with respect to the whole network. Although the notion of centrality
in graph theory dates back to the mathematician Camille Jordan (1869), centrality measures 
were expanded, refined and applied to a great extent for the first time in social sciences  
\cite{wasserman,scott, sabidussi,friedkin,SocNet_BE06}, and today they play a fundamental role
in studies involving a large variety  of  complex networks across many fields.
Probably the most frequently used centrality measure is betweenness centrality (BC) 
\cite{anthonisse,freeman77,SocNet_F79,SocNet_BE06,SocNet_B08,
WhiteBorgatti94,SocNet_B05},  introduced by Anthonisse \cite{anthonisse} and Freeman 
\cite{freeman77} defined as the fraction of all network geodesics (shortest paths) passing through 
a node (edge or subgraph).
Since transport tends to minimize the cost/time of the route from source to destination, 
it expectedly happens along geodesics,  and therefore centrality measures 
are typically defined as a function of these, however generalizations to arbitrary distributions of
transport paths have also been introduced and studied \cite{PRE_SCTS07,JACM_DEP10}. 
Geodesics are important for structural connectivity as well: removing nodes (edges) 
with high BC, one obtains a rapid increase in  diameter, and 
eventually the structural breakup of the  graph. 

In general, centrality measures are defined in the context of the assumptions 
(sometimes made implicitly) regarding the type of network flow \cite{SocNet_B05}. 
These are assumptions regarding the {\em nature of the paths} such as being 
shortest, or arbitrary length paths, weighted/valued paths, walks (repeated nodes and edges) 
\cite{BollobasBook}  etc.; and the 
{\em nature of the flow}, such as transport of indivisible units (packets), or 
spreading/broadcasting processes (infection, information).  
Besides betweenness centrality, many other centrality measures have been introduced
\cite{SocNet_BE06}, depending on the context in which network flows are considered; 
for a partial compilation see the paper by Brandes \cite{SocNet_B08}, here we only review 
a limited list. In particular, {\em stress} centrality 
\cite{shimbel,Perer2006,Lammer2006}, simply counts the number of all-pair shortest paths 
passing through a node (edge) without taking into account the degeneracy
of the geodesics (there can be several geodesics running between the same pair of nodes). 
{\em Closeness} centrality \cite{sabidussi,SocNet_F79,JGraphAlgAppl_EW04} and its 
variants are simple functions of the mean geodesic distance (hop-count) of a node 
from all other nodes. 
{\em Load} centrality  \cite{PhysRevLett_GKK01,PhysRevE_N01,SocNet_B08} is 
generated by the total amount of load passing through a node when unit commodities 
are passed between all source-destination pairs using an algorithm in which the 
commodity packet is equally divided amongst the neighbors of a node that are at 
the same geodesic distance from the destination. 
{\em Group} betweeness centrality \cite{EverettBorgatti99,Puzis2007} computes the 
betweenness associated with a set of nodes restricted to all-pair geodesics that traverse
at least one of the nodes in the group. 
{\em Ego} network betweenness \cite{SocNet_EB05} is a local betweenness measure 
computed only from the immediate neighborhood of a node (ego). 
{\em Eigenvector} centrality 
\cite{JMathSoc_B72,SocNet_B07} represents a positive
score associated to a node, proportional to the sum of the scores of the node's
neighbors, solved consistently across the graph. The corresponding score vector is
the eigenvector associated with the largest eigenvalue of the adjacency matrix.
{\em Random walk} centrality \cite{PhysRevLett_NR04,NewmanBCrandwalk} 
is a measure of the accessibility of a node via random walks in the network.
Other centrality measures include {\em information} centrality of Stephenson and Zelen
\cite{SocNet_SZ89}; and {\em induced endogenous and exogenous} centrality by 
Everett and Borgatti \cite{SocNet_EB10}.

{\em Bounded-distance} betweenness was introduced
by Borgatti and Everett \cite{SocNet_BE06} as betweenness centrality resulting from 
all-pair shortest paths not longer than a given length (hop-count). It is this measure 
that we expand and investigate in detail in the present paper. A condensed version for
unweighted networks has been presented in Ref.  \cite{PRLcikkunk}.  Since we are also 
generalizing the measure and the corresponding algorithm to weighted (valued) 
networks, we are referring to it as {\em range-limited} centrality. Note that range-limitation
can be imposed on all centrality measures that depend on paths, and therefore the
analysis and algorithm presented here can be extended to all these centrality measures. 

Centrality measures have received numerous applications in several areas. 
In social sciences they have been extensively used to quantify the position of individuals 
with respect to the rest of the network in various social network 
data sets \cite{wasserman,SocNet_B05}. 
In physics and computer science they have seen widespread applications 
related to routing algorithms in packet switched 
communication networks and transport problems in general \cite{PhysRevLett_GKK01,
PhysRevLett_ADG01, PhysRevLett_GDVCA02,
PhysRevE_YZHFW06, Danila, DanilaPRE2006, DanilaCHAOS2007,PRE_SCTS07}. The connection of generalized betweenness 
centrality based on arbitrary path distributions (not just shortest) to routing that minimizes congestion 
has been investigated by Sreenivasan {\em et al} \cite{PRE_SCTS07} using minimum 
sparsity vertex separators. This makes a direct connection to max-flow min-cut theorems 
of multicommodity flows, extensively studied in the computer science literature \cite{JACM_LR99,Book_Vazirani_03}. Other works that use essentially edge betweenness 
type quantities to quantify congestion in Internet-like graphs include Refs
\cite{SIGMETRICS_GMS03,PODC_ACKS03}. Dall'Asta et.al. connect node and edge
detection probabilities in traceroute-based sampling of networks to their betweenness 
centrality values \cite{vespignani-traceroute,PRE-int-explor}.
Other applications include detection of network vulnerabilities in face of attacks 
\cite{holme}, cascading failures \cite{PhysRevE_ML02,motter,vesp09} or epidemics \cite{vespignani}, 
all involving betweenness-related calculations.  

An important extension of centrality is to weighted, or valued networks 
\cite{SocNet_FBW91,PhysRevE_N01,JMathSoc_B01,
PNAS_BBPV04,PhysRevE_WHV08,SocNet_OAS10}. 
In this case the edges (and also the nodes) carry an associated weight, which 
may represent a measure of social relationship in social networks \cite{AmJSoc_G73},
channel capacity in the case of communication networks, transport capacity (e.g., nr of lanes)
in roadway networks or seats on flights \cite{NatPhys_CPV07}.

From a theory point of view, there have been fewer results, as producing analytic 
expressions for centralities in networks is difficult in general. However, for scale-free trees,
Szab\'o et.al. 
\cite{PhysRevE_SzAK02} developed a mean-field approach for computing node betweenness, 
which later was made rigorous
by Bollob\'as and Riordan \cite{PhysRevE_BR04}. Fekete et.al. provide a calculation of the 
distribution of edge betweeness on scale-free trees conditional on node in-degrees
 \cite{PhysRevE_FVK06}, and Kitsak et.al. \cite{PhysRevE_KHPRPS07} have derived 
 scaling results on betweenness centrality for fractal and non-fractal scale-free networks. 

Unfortunately, computation of betweenness can be costly  (${\cal O}(N M)$, 
where $N$ is the number of nodes and $M$ is the number of edges, thus 
${\cal O}(N^3)$ worst case) \cite{JMathSoc_B01,PhysRevE_N01,SocNet_B08,
johnson,floyd, warshall}, especially for large networks with millions 
of nodes, hence approximation methods are needed. 
Existing approximations \cite{brandes-approx,geisberger-approx,NewmanBCrandwalk}, however, 
are sampling based, and not well controlled. Additionally, transport in real networks 
does not occur with uniform probability between arbitrary pairs of nodes, as transport incurs
a cost, and therefore shorter-range transport is expectedly more frequent than long-range. 
Accordingly, the {\em usage} of network paths is non-uniform, which should be taken 
into account if we want to connect centrality properties with real transport. 
\begin{figure*}[htbp] \begin{center}
\includegraphics[width=0.9\textwidth]{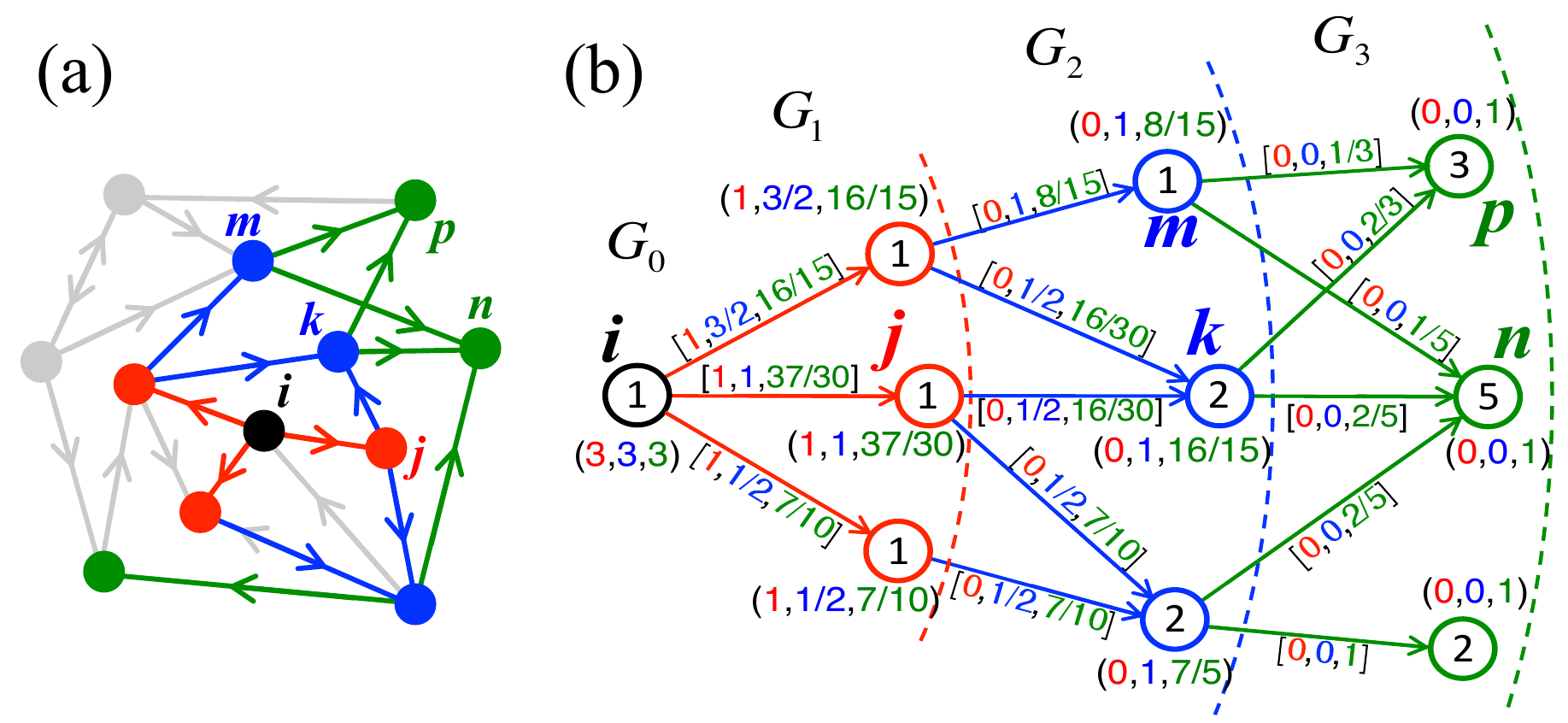}
\caption{a) Consecutive shells of the $\bm{C}_3$ subgraph of node $i$ (black) are colored 
red, blue, green. Grey elements are not part of the subgraph. b) The $(x,y,z)$ near a node $j$
are the $b_{\ell}(i|j)$ values for $\ell=1$, $\ell=2$ and $\ell=3$. The $[x,y,z]$ on an edge $(j,k)$ give the $b_{\ell}(i|j,k)$ values for $\ell=1$, $\ell=2$ and $\ell=3$ . Given a node $j$, the number inside its circle 
is the total number of shortest paths $\sigma_{ij}$ to $j$ from $i$. Colors indicate quantities based on $\ell =1$ (red), $\ell =2$ (blue), and $\ell =3$ (green).} \label{fig1} 
\vspace*{-0.5cm} \end{center} \end{figure*}
In order to address some of the limitations of existing centrality measures,
we recently focused on range-limited centrality \cite{PRLcikkunk}. We have shown
 that when geodesics are restricted to a maximum length $L$, the 
corresponding range-limited $L$-betweenness for large graphs 
assumes a characteristic scaling form as function of $L$. This scaling can then be 
used to predict the betweenness distribution in the (difficult to attain)
diameter limit, and with good approximation, to predict the ranking of nodes/edges 
by betweenness, saving considerable computational costs. 
Additionally, the range-limited method generates $l$-betweenness 
values {\em for all} nodes and edges and {\em for all} $1 \leq l \leq L$, providing
systematic information on geodesics on all length-scales. 

In this paper we give a detailed derivation of the algorithm and the analytical approximations
presented in \cite{PRLcikkunk} and we  demonstrate the efficiency of the method
on a social network (SocNet) inferred from mobile phone trace-logs \cite{Barabasi-mobility}.
This network has  a giant cluster with $N=5,568,785$ nodes and $M=26,822,764$ directed edges.
The diameter of the underlying undirected network is approximately $D\simeq 26$ and the calculation
of the traditional (diameter-range based) BC values (using Brandes' algorithm) on this network
took $5$ days on $562$ computers. 

In addition, we present the derivations for an algorithm that  efficiently computes 
range-limited centralities on {\em weighted} networks. We then apply these concepts and algorithms to the network vulnerability backbone detection problem, and show the differences
between the backbones obtained with both hop-count based centralities and weighted
centralities. 

The paper is organized as
follows. Section \ref{rlcnw} introduces the notations and provides the algorithm for unweighted
graphs; section \ref{scaa} gives an analytical treatment that derives the existence of a scaling
behavior for centrality measures in large graphs; it gives a method on how to estimate the largest
typical node-to-node distance (a lower-bound to the diameter); discusses the complexity of the
algorithm and the fast freezing phenomenon of ranking by betweenness of nodes and edges. Section
\ref{rlcsn} illustrates the power of the range-limited approach (by showing how well can
one predict betwenness centralities and ranking of individual nodes and edges) 
using the social-network data described above. Section \ref{rlcwn} describes the algorithm 
for weighted graphs and section \ref{vbb} uses the range-limited betwenness measure to
define a vulnerability backbone for networks and illustrates the differences in identification of
the backbone obtained with and without weights on the links.

\section{Range-limited centrality for non-weighted graphs} \label{rlcnw}

\subsection{Definitions and notations}

Let us  consider a directed simple graph   $\bm{G}(V,\bm{E})$, which consists of a set $V$ of 
vertices (or nodes) and a set $E\subseteq V \times V$ of directed edges (or links). We
will denote by $(v_i,v_j)\in E$ an edge directed from node  $v_i \in V$ to node $v_j\in V$.
The graph has  $N$  nodes and $M \leq N(N-1)$ edges. The algorithm below can easily be modified for undirected graphs, we will not treat that case separately.
A directed path $\omega_{mn}$ from some node $m$ to a node $n$ is defined 
as an ordered sequence of nodes  and links 
$\omega_{mn} = \left\{ m, (m,v_1),v_1,(v_1,v_2),v_2,....v_l,(v_l,n),n\right\}$ without repeated nodes.  
The ``distance'' $d(m,n)$ is the length of the shortest {\em directed} 
path going from node $m$ to node  $n$. We give a definition of distance 
(path weight) for weighted networks in Section \ref{rlcwn}. In non-weighted networks the 
directed path length is simply the number of edges (``hop-count") 
along the directed path from $m$ to $n$.  There can be multiple shortest paths 
(same length), and we will denote by $\sigma_{mn}$ the total number   of 
shortest directed paths from node $m$ to $n$. $\sigma_{mn}(i)$ will represent 
the number of shortest paths from node $m$ to node $n$ {\it going through} node $i$.  As convention
we set 
\begin{equation}
\sigma_{mn}(m) = \sigma_{mn}(n) = \sigma_{mn},\;\; 
\sigma_{mm}(i) = \delta_{i,m}\;.\label{conv}
\end{equation}

The total number of all-pair shortest paths running through a node 
$i$ is called the stress centrality (SC) of node $i$,
$S(i)=\sum_{m, n \in V}\sigma_{mn}(i)$.
 Betweenness centrality (BC) \cite{freeman77, anthonisse,SocNet_BE06,SocNet_B08,JMathSoc_B01} normalizes the number of paths 
 through a node by the total number of paths ($\sigma_{mn}$) 
 for a given source-destination pair $(m,n)$:
\begin{equation}
B(i)=\sum_{m, n \in V}\frac{\sigma_{mn}(i)}{\sigma_{mn}}.
\end{equation}
Similar quantities can be 
defined for an edge $(j,k) \in \bm{E}$:  
\begin{equation}
B(j,k)=\!\!\!\!\sum_{m, n \in V}\!\!\!\frac{\sigma_{mn}(j,k)}{\sigma_{mn}}\;.
\end{equation}

In order to define range-limited betweenness centralities, let  $b_l(j)$ denote the 
BC of a  node $j$ for all-pair shortest directed 
paths of {\em fixed, exact} length $l$. Then 
\begin{equation}
B_L(j) = \sum_{l=1}^L b_l(j)  \label{BL}
\end{equation}
represents the betweenness centrality
obtained from paths {\em not longer} than 
$L$. For edges, we introduce $b_l(j,k)$  and $B_L(j,k)$ 
using the same definitions. 
For simplicity, here we include the start- and end-points of the paths in the centrality 
measures,  however, our algorithm can easily be changed to exclude them, as
described later. 

Similar to other algorithms, our method first calculates these BCs
for a node $j$ (or edge $(j,k)$)
from shortest directed paths all emanating from a ``root'' node $i$, 
then it sums the obtained values for all $i \in V$  to get the final centralities for 
node $j$ (or edge $(j,k)$). This can be done because the set of 
{\em all} shortest paths can be uniquely decomposed into subsets of 
shortest paths distinguished by their starting node.
Thus it makes sense to perform a shell decomposition of the graph 
around a root node $i$. Let us denote by 
$\bm{C}_L(i)$  the $L$-range subgraph 
of node $i$ containing all nodes which can be reached  
in at most $L$ steps
from $i$ (Fig.~\ref{fig1}a). Only links which are part of the shortest paths starting 
from the root $i$ to these nodes are 
included in $\bm{C}_L$. We decompose $\bm{C}_L$ into shells $G_l(i)$ containing  
all the nodes at shortest path distance $l$  from the root, and all incoming edges
from shell $l-1$, Fig.~\ref{fig1}b). The root $i$ itself is considered to be shell $0$ 
($G_0(i) = \{i\}$). Let 
\begin{eqnarray}
b_{l}^{r}(i|k) = \!\!\!\sum_{n\in G_l(i)}\!\!\!\frac{\sigma_{in}(k)}{\sigma_{in}}\;,\,\;\;  
b_{l}^{r}(i|j,k) = \!\!\!\sum_{n\in G_l(i)}\!\!\!\frac{\sigma_{in}(j,k)}{\sigma_{in}}   \label{blr}
\end{eqnarray}
denote the fixed-$l$ betweenness centrality 
of node $k$, and edge $(j,k)$, respectively, based only on shortest paths 
all starting from the root $i$.
Here $r$ is {\em not} an independent variable:
given $i$ and $k$ (or $(j,k)$), $r$ is the radius of  shell $G_r(i)$ 
containing $k$ (or $(j,k)$), that is $k \in G_r(i)$ and $(j,k) \in G_r(i)$. 
Note that $\sigma_{in}(k) = 0$ (or $\sigma_{in}(j,k) = 0$) if 
$k$ (or $(j,k)$) do not belong to at least one shortest path from $i$ to $n$, and thus there is
no contribution from those points $n$ from the $l$-th shell. The condition
for $k$ (or $(j,k)$) to belong to at least one shortest path from $i$ to $n$ can alternatively
be written in the case of (\ref{blr}) as $d(k,n) = l-r$ a notation, which we will use later.

For simplicity of writing, we refer to the
fixed-$l$ betweenness centralities (the $b_l$-s) as ``$l$-BCs" and to the
cumulative betweenness centralities 
(the $B_L$-s obtained from summing the $l$-BCs, see (\ref{BL})) as $[L]$-BCs.

\subsection{The range-limited betweenness centrality algorithm}

While the basics of our algorithm are similar to  Brandes' \cite{JMathSoc_B01,SocNet_B08}, 
we derive recursions that simultaneously compute 
the $[l]$-BCs for {\em all} nodes and {\em all} edges and for all values $l=1,\ldots,L$. 
The algorithm thus generates detailed and systematic information (an $L$-component
{\em vector} for every node and every edge)
about shortest paths on all length-scales and thus, providing a tool for 
multiscale network analysis.

First we give the algorithm, then we derive the specific recursions used in it.
For the root node $i$ we set the initial condition: $\sigma_{ii}=1$. 
For other nodes, $k\neq i$, we set $\sigma_{ik}=0$.
The following steps are repeated for every $l=1,\dots ,L$:
\begin{enumerate}
\item[1.] Build $G_l(i)$, using breadth-first search. 
\item[2.] Calculate $\sigma_{ik}$  for  all nodes $k\in G_l(i)$, using:
\begin{equation}
\sigma_{ik}= \!\!\!\sum_{{j\in G_{l-1}(i) \atop (j,k)\in G_{l}(i)}}  \!\!\! \sigma_{ij}\;,\label{sigma}
\end{equation}	 
and set 
\begin{equation}
b_l^l(i|k)=1.  \label{bllequal1}
\end{equation}	 
\item[3.]  Proceeding {\em backwards}, through $r=l-1,\ldots,1,0$:

a) Calculate the $l$-BCs  of links $(j,k)\in G_{r+1}(i)$ (thus
$j \in G_r(i)$, $k \in G_{r+1}(i)$) recursively:
\begin{equation}
b_{l}^{r+1}(i|j,k) = b_{l}^{r+1}(i|k) \frac{\sigma_{ij}}{\sigma_{ik}}\;, \label{3a}
\end{equation}
b) and of nodes $j \in G_r(i)$ using (\ref{3a}) and:
\begin{equation}
b_{l}^{r}(i|j) = \!\!\!\sum_{{k\in G_{r+1}(i) \atop (j,k)\in G_{r+1}(i)}} \!\!\! b_{l}^{r+1}(i|j,k)\;.   \label{3b}
\end{equation}
\item[4.] Finally, return to step 1) until the last shell $G_L(i)$ is reached. 
\end{enumerate}
In the end, the cumulative  $[l]$-BCs, that is the $B_l$-s can be calculated using (\ref{BL}). Fig. 1
shows a concrete example. The subgraph of node $i$ has three layers. 
Each layer $G_l(i)$ and the corresponding $l$-BCs are marked with 
different colors: $l=1$ (red), $l=2$ (blue), and $l=3$ (green).
As described above, the first step creates the next layer $G_l(i)$, 
then in step 2., for every 
node $k\in G_l(i)$ we calculate the total number 
of shortest paths $\sigma_{ik}$ from the root to node $k$. These are indicated by  numbers
within the circles representing the nodes in Fig.~\ref{fig1}
(e.g., $\sigma_{ij}=1$,  $\sigma_{ik}=2$, $\sigma_{in}=5$). As given by (\ref{sigma}), $\sigma_{ik}$  is calculated by summing the number of shortest paths that end in the predecessors of node $k$ located in 
$G_{l-1}(i)$. For example node $p \in G_3(i)$ in Fig.~\ref{fig1} is connected to nodes $k$  and
$m$ in shell $G_2(i)$, and thus: $\sigma_{ip}=\sigma_{ik}+\sigma_{im}=2+1=3$.

Eq. (\ref{bllequal1}) states that the $l$-BC of nodes located in $G_l(i)$ is always $1$. 
This follows from  Eq. (\ref{blr}) for $r=l$  and using the convention $\sigma_{ik}(k)=\sigma_{ik}$. 
Knowing these values,  we proceed backwards (step 3.) and calculate the 
$l$-BCs of all edges and nodes in all the previous layers. 
Recursion (\ref{3a}) is obtained from a well known recursion for 
shortest paths. If $k$ (or $(j,k)$) belongs to at least one shortest path going from
$i$ to $n$, then  $\sigma_{in}(k)=\sigma_{ik} \sigma_{kn}$ and 
$\sigma_{in}(j,k)=\sigma_{ij} \sigma_{kn}$. 
Inserting these in Eq. (\ref{blr}) for $r \mapsto r+1$ we obtain:
\begin{eqnarray}
b_{l}^{r+1}(i|k) &=& \sigma_{ik}\!\!\!\!\sum_{{n\in G_l(i) \atop d(k,n)=l-r-1}}\!\!\!\!\frac{\sigma_{kn}}{\sigma_{in}}\\
 b_{l}^{r+1}(i|j,k) &=& \sigma_{ij} \!\!\!\!\sum_{{n\in G_l(i)  \atop d(k,n)=l-r-1 }}\!\!\!\!\frac{\sigma_{kn}}{\sigma_{in}}   
\end{eqnarray}
where $d(k,n)=l-r-1$ expresses the condition that the sum is restricted to those $n$ from $G_l(i)$,
which have at least one shortest path (from $i$), going through  $k$ or $(j,k)$. 
Dividing these equations we obtain (\ref{3a}). For e.g., in 
Fig~\ref{fig1}: $b_3^3(i|k,n)=b_3^3(i|n)\sigma_{ik} / \sigma_{in} =1\times (2/5)=2/5$.

Having determined the $l$-BCs of  all edges in layer $G_{r+1}(i)$, we 
can now compute the $l$-BC of a given node in $G_r(i)$ by summing 
the $l$-BCs of its outgoing links, that is using (\ref{3b}) (e.g., on 
Fig~\ref{fig1}: $b_3^2(i|k)=b_3^3(i|k,p)+b_3^3(i|k,n)=(2/3)+(2/5)=16/15$).

This  algorithm can  be easily modified to compute other centrality measures. 
For example, to compute all the range-limited stress centralities, we have to replace 
 Eq. (\ref{bllequal1}) with: $s_l^l(i|j)=\sigma_{ij}$. All other recursions will 
 have exactly the same form, we just need to
 replace the $l$-BCs ($b_l^r(i|j)$, $b_l^r(i|j,k)$)  with the $l$-SCs ($s_l^r(i|j)$, 
 $s_l^r(i|j,k)$).

If we want to exclude start- and end-points when computing BCs or SCs,  
we first let the above algorithm finish, then we do the following steps:
a) set the $l$-BC of the root node $i$ to $0$, $b_l^0(i|i)=0$ for all $l=1,\dots,L$, 
and b) for every node $k\in G_l(i)$ reset  $b_l^l(i|k)=0$, for all $l=1,\dots,L$, 
(for e.g., on Fig~\ref{fig1} $k$ is in the second shell, $G_2(i)$, so its 
$2$-BC will become $0$ instead of $1$). Then via (\ref{BL}), the $[l]$-BCs and
the corresponding $[l]$-SCs are easily obtained.

\section{Centrality Scaling - Analytical approximations} \label{scaa}

In \cite{PRLcikkunk} we have shown that the $[l]$-BC 
obeys a scaling behavior as function of $l$. This was found to hold
for all sufficiently large random networks that we studied
(Erd\H{o}s-R\'enyi (ER), Barab\'asi-Albert (BA) scale-free, 
Random Geometric Graphs (RGG), etc.) including the social network inferred 
from mobile phone trace-log data (SocNet) \cite{PNAS_OSHSLKKB07}. Here we detail the analytical 
arguments that indeed show that the
existence of this scaling behavior for large networks  is a general property, by 
exploiting the scaling of shell sizes. The scaling of shell sizes was already studied previously, for e.g., in random graphs
 with arbitrary degree distributions \cite{newman_randomgraphs, shell_structure}.  
 For simplicity of the notations, we only show the derivations for undirected graphs.
 
 \subsection{Betweenness of individual nodes}
 
Let us define  $\langle\cdot\rangle$ as an average over all root 
nodes $i$ in the graph, and denote by $z_{l}(i)$   the number of nodes on 
shell $G_{l}(i)$. We define the branching factor as:
\begin{equation}
  \alpha_{l}=\langle z_{l+1}\rangle /\langle z_l \rangle\;,
  \end{equation}
and model the growth of shell sizes as a branching process 
  \begin{equation}
z_{l+1}(i)=z_{l}(i)\alpha_{l}\big[1+\epsilon_{l}(i)\big]\;.   \label{zbranching}
\end{equation}
Here   $\epsilon_{l}(i)$ is a {\em per-node}, shell occupancy 
noise term, encoding the relative deviations, or fluctuations from the ($i$-independent) 
functional form of $\alpha_l$. Typically, $|\epsilon_l|\ll1$, it obeys
$\langle \epsilon_{l}(i) \rangle = 0$ and 
$\langle \epsilon_{l}(i) \epsilon_{m}(j)\rangle = 2A_l \delta_{l,m} \delta_{i,j}$, with 
$A_l$ decreasing with $l$. 
In undirected graphs if $i\in G_m(j)$ then it implies that $j\in G_m(i)$, and 
vice-versa. Hence, in this case:
 \begin{eqnarray}
 b_{l+1}(j)=\frac{1}{2}\sum_{i\in V}b_{l+1}(i|j) = 
 \frac{1}{2}\sum_{m=0}^{l+1}\sum_{i\in G_m(j)} b_{l+1}^m(i|j)
\end{eqnarray}
 The $1/2$ factor  comes from the fact that  any given path  will 
 be included twice in the sum (once in both directions). In case of $m=0$ 
 the only node in $G_0(j)$ is $j$ itself,  and  the inner sum is equal with 
 $b_{l+1}^0(j|j)$. Due to convention (\ref{conv}) $\sigma_{jn}(j)=\sigma_{jn}$
 and hence from (\ref{blr}) we obtain
 $b_{l+1}^0(j|j)=\sum_{n\in G_{l+1}(j)} \sigma_{jn}(j)/\sigma_{jn} = z_{l+1}(j)$.  
 For $m=l+1$,  $b_{l+1}^{l+1}(i|j)=1$ (see Eq. (\ref{bllequal1})) and 
 the inner sum is again $z_{l+1}(j)$. Thus we can write:
 \begin{eqnarray}
b_{l+1}(j) &=& z_{l+1}(j)+ \frac{1}{2}\sum_{m=1}^{l}\sum_{i\in G_m(j)} 
b_{l+1}^m(i|j) \equiv \nonumber \\
&\equiv& z_{l+1}(j)+ \frac{1}{2} u_{l+1}(j),    \label{buszerint}
\end{eqnarray}
Note that the  number of terms in the inner sum $\sum_{i\in G_m(j)} b_{l+1}^m(i|j)$ 
is $z_m(j)$, which is rapidly increasing with $m$, and thus  is expected 
to have a weak dependence on $j$. Accordingly, we make the approximation:
\begin{equation} 
 u_{l+1}(j) \simeq \sum_{m=1}^{l}\sum_{i\in G_m(j)} v^m_{l+1}(i),  \label{ulplus1}
\end{equation} 
where we replaced $b_{l+1}^m(i|j)$ by $v^m_{l+1}(i)$, which is an {\em average} 
$(l+1)$-BC computed over the shell of radius
$m$, {\em centered  on node} $i$ : 
\begin{equation}
v^m_{l+1}(i)=\frac{\sum_{k\in G_{m}(i)}  b_{l+1}^m(i|k)}{z_{m}(i)}.
\end{equation}
However, the sum of ($l+1$)-BCs in any $m\leq l+1$ layer
is equal with the number of nodes in shell $G_{l+1}$:
$\sum_{k\in G_m(i)}b_{l+1}^m(i|k)=z_{l+1}(i)$. We can convince 
ourselves about this last statement by using (\ref{blr}) and observing that
$\sum_{k\in G_m(i)} \sigma_{in}(k) = \sigma_{in}$ as all paths from $i$ 
to $n$ ($n \in G_{l+1}(i)$) must ``pierce" every shell $m \leq l+1$ in between.
Fig.~\ref{fig1} shows an example: there are $3$ nodes in $G_3$ and the sum 
of $3$-betweenness values (green) in layer $G_2$ is $(7/5)+(16/15)+(8/15)=3$. 
Therefore, we may write:
\begin{equation}
v^m_{l+1}(i) \simeq \frac{z_{l+1}(i)}{z_{m}(i)}=\frac{z_{l}(i)\alpha_{l}
\big[1+\epsilon_{l}(i)\big]}{z_m(i)}\;, \label{vapprox}
\end{equation}
where we used the recursion defined above for $z_{l+1}(i)$ 
as a branching process (\ref{zbranching}).
Inserting this in (\ref{ulplus1}) we obtain:
\begin{eqnarray}
&u_{l+1}(j)& \simeq \alpha_l \sum_{m=1}^{l}  
\sum_{i\in G_m(j)} \frac{z_{l}(i)\big[1+\epsilon_{l}(i)\big]}{z_{m}(i)} 
\simeq \nonumber \\
&\simeq& \alpha_l \sum_{m=1}^{l}  \sum_{i\in G_m(j)} 
\frac{z_{l}(i)}{z_{m}(i)}  \simeq \nonumber \\
&\simeq& \alpha_l  \left[ z_{l}(j)+ \sum_{m=1}^{l-1}  
\sum_{i\in G_m(j)} \frac{z_{l}(i)}{z_{m}(i)}\right]   \label{ukifejtve}
\end{eqnarray}
where we neglected the small noise term due to 
the large number of terms in the inner sum, and we 
used the fact that for $m=l$ the leading term of the 
inner sum is just $z_l(j)$.
From Eqs. (\ref{ulplus1}) and (\ref{vapprox}), however, the 
double sum in (\ref{ukifejtve}) equals $u_l(j)$ and we obtain the
following recursion:
\begin{equation}
u_{l+1}(j) \simeq \alpha_l \big[ z_l(j)+u_l(j)\big]. \label{urekurzio}
\end{equation}
Eqs (\ref{zbranching}), (\ref{buszerint}) and (\ref{urekurzio}) 
lead to a recursion  for $b_{l+1}(j)$:
\begin{equation}
b_{l+1}(j)\simeq \alpha_l [b_l(j)+ z_l(j)/2 +z_l(j)\epsilon_{l}(j)],
\end{equation}
which can be iterated down to $l=1$, where 
$b_1(j)=z_1(j)=k_j$ is the degree of $j$:
\begin{equation}
b_{l}(j) \simeq \beta_{l}\; k_j \;e^{\xi_l(j)}\;, \label{nice}
\end{equation}
with 
\begin{eqnarray}
\beta_{l} &=& \frac{l+1}{2} \prod_{m=1}^{l-1}\alpha_m 
=  \frac{l+1}{2} \frac{\langle z_{l}\rangle}{ \langle k \rangle}\;, \label{beta} \\
\xi_{l}(j) &=&\sum_{n=1}^{l-1}\!\frac{l+1-n}{l+1}\;\epsilon_{n}(j)\;. \label{xi}
\end{eqnarray}
In many networks,  the average shell-size $\langle z_{l}\rangle$ grows 
exponentially with the shell-`radius' $l$ 
(for e.g., ER, BA, SocNet), implying a constant average branching factor larger
than one:
\begin{equation}
\alpha_l \simeq \alpha = \frac{\langle z_2 \rangle}{\langle k \rangle} > 1\;.  \label{abf}
\end{equation}
The exponential growth holds until  $l$ reaches the typical largest shortest 
path distance $L^*$, beyond which finite-size effects appear. 
Accordingly, $\beta_l\sim \alpha^l$ 
and $b_l$ grows exponentially with $l$. 
In this case,  since $b_l$ is rapidly increasing with $l$, the cumulative 
$B_L(j)= \sum_{l=1}^{L} b_l(j)$ will be dominated
by $b_L$, and thus $B_L$ obeys the same exponential 
scaling as $b_l$, confirmed by numerical 
simulations (Fig. 3c in \cite{PRLcikkunk} shows this scaling for SocNet). 

However, not all large networks have exponentially growing shell-sizes. 
For example, in spatially embedded networks without shortcuts such as 
random geometric graphs, roadways, etc., average shell-size grows as 
a {\em power law} $\langle z_l \rangle \sim  l^{d-1}$, where  $d$ 
is the embedding dimension of the metric space. In this case $\beta_{l} \sim l^{d}$ and  $b_l(j)
\sim l^{d}$ and $B_L \sim L^{d+1}$. Fig 3d in \cite{PRLcikkunk} shows this scaling for RGG graphs
embedded in $d=2$ dimensions.

\subsection{Distribution of $l$-betweenness centrality}

Eq (\ref{nice}) allows to relate the statistics of fixed-$l$ betweenness to the 
statistics of shell occupancies for networks that are uncorrelated, or short-range 
correlated.  Since the noise term (obtained from {\em per-node} occupancy deviations 
on a shell) is independent on the root's  degree in this case, 
the distribution of fixed-$l$ betweenness can be expressed as:
\begin{eqnarray}
\rho_{l}(b) &=&\langle \delta\left( b_l(j)-b\right) \rangle  = \nonumber \\
&=& 
\int_{-\infty}^{\infty}\!\!\!\!\!\! d\xi \int_{1}^{N-1} \!\!\!\!\!\!dk \;\delta \left(\beta_l k e^{\xi} - b \right)P(k)
\Phi_l(\xi)\;. \label{derivrho}
\end{eqnarray}
where $\delta(x)$ is the Dirac-delta function,  $P(k)$ is the degree distribution and $\Phi_l(\xi)$ is the distribution for the noise 
$\xi_l(j)$, peaked at $\xi = 0$, with fast decaying tails and $\Phi_1(x) = \delta(x)$.  Performing
the integral over the noise $\xi$, one obtains the distribution for $l$-BC, in form of a 
convolution:
\begin{equation}
\rho_{l}(b) = 
\frac{1}{b}\int_1^{N-1}\!\!\!\!\!\!\!dk \;P(k) \Phi_l(\ln b - \ln \beta_l - \ln k)\;. \label{bd}
\end{equation}
From (\ref{bd}) follows that 
the natural scaling variable for betweenness distribution is $u = \ln b - \ln \beta_l$. 
The noise distribution $\Phi_l$ (for $l > 1$) may introduce an extra $l$-dependence
through its width $\sigma_l$, which can be accounted for via the rescaling 
$u \mapsto u/\sigma_l$, $\rho_l \mapsto \rho_l \sigma_l$, thus collapsing the distributions 
for different $l$-values onto the same functional form, directly supporting our 
numerical observations presented in Ref \cite{PRLcikkunk}.
As  $\Phi_l$ is typically sharply peaked around 0, the most significant contribution 
to the integral (\ref{bd}) for a given $b$ comes from degrees $k \simeq b/\beta_l$. 
Since $k \geq 1$, we have a rapid decay of $\rho_l(b)$ in the range $b < \beta_l$,
a maximum at $\overline{b} = \beta_l \overline{k}$ where $\overline{k}$ is the degree at
which $P(k)$ is maximum, and a sharp decay for $b > (N-1) \beta_l$.

\subsection{Estimating the average node-to-node distance in large networks.} 
\label{L*}

\begin{figure}[htbp] \begin{center}
\includegraphics[width=0.4\textwidth]{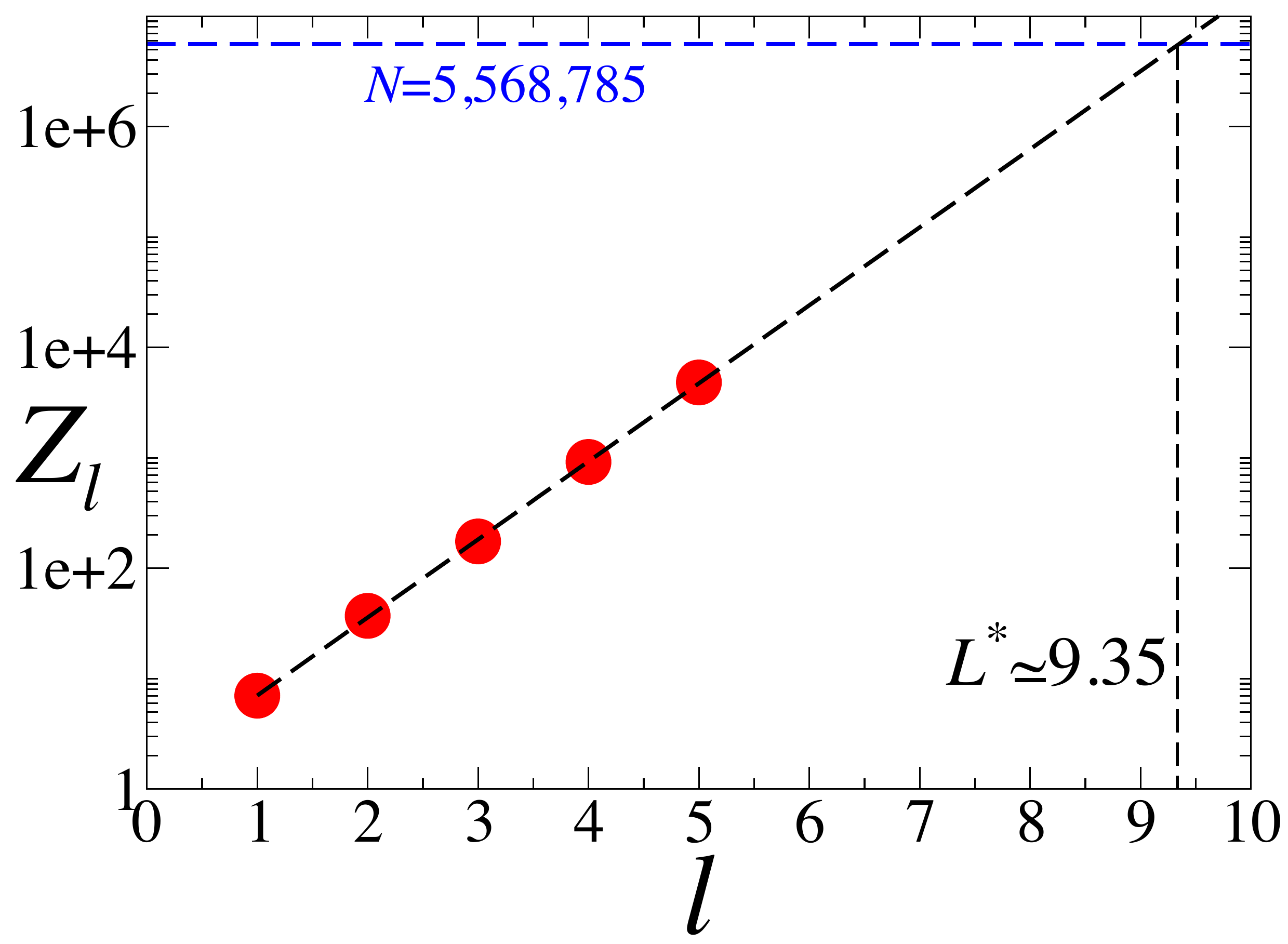}
\caption{In the SocNet the sum, $Z_l$,  of the average shell sizes 
grows exponentially as function of $l$. Extrapolating, we can predict
that it reaches the $N=5,568,785$ mark at $L^*\simeq9.35$.}\label{Lcsillag} 
\vspace*{-0.5cm} \end{center}
 \end{figure}

The scaling law on its own does not 
provide information about the typical largest  node-to-node distance, which is always 
a manifestation of the
finiteness of the graph. However, knowing the size of the network in terms of the
number of nodes $N$, one can exploit our formulas to find the average largest node-to-node
distance as the radius $L^*$ of the {\em typical 
largest shell} beyond which finite-size effects become strong, that is where 
network edge effects appear. 
This can be estimated as the point where the sum of the average shell sizes 
reaches $N$. Hence:
\begin{equation}
Z_{L^*}=\sum_{l=1}^{L^*} \langle z_l \rangle = \sum_{l=1}^{L^*} 
\frac{2}{l+1} \langle k \rangle\beta_l = N \;,\label{fit}
\end{equation}
providing an {\em implicit equation} for $L^*$.   
The $\beta_l$-s
are determined numerically for $l=1,2,3,\ldots$ and a corresponding functional form
fitting its scaling with $l$ 
can be extrapolated for larger $l$ values up to $L^*$, when the sum in (\ref{fit}) hits $N$.
For our social network data one obtains $L^*\simeq 9.35$ (Fig.~\ref{Lcsillag}). 
Here $L^*$ is not necessarily an integer, because it is obtained from the scaling 
behavior of the average shell sizes, and  represents  the {\it typical} radius of the largest shell.
 Expression (\ref{fit}) can be easily specialized for the two classes of 
networks discussed above namely, for  those having exponential average shell-size growth 
 $\langle z_l \rangle \sim  \langle k \rangle \alpha^{l-1}$ and for those having a power-law 
 average shell-size growth as $\langle z_l \rangle \sim\langle k \rangle l^{d-1}$. 
 For the exponential growth case we obtain:
 \begin{equation}
 L^* = \frac{1}{\ln{\alpha}}\ln{\left(1+\frac{\alpha - 1}{\langle k \rangle} N\right)}\;, \label{Lexp}
 \end{equation}
resulting in the $L^* \sim \ln N$ behavior for large $N$. 
 
 For the power-law growth case there is no easily invertible expression for the sum, however,
 if we replace the summation with an integral, we find the approximate 
  \begin{equation}
 L^* \simeq \left( 1+ \frac{d}{\langle k \rangle} N\right)^{1/d}  \label{Lpow}
 \end{equation}
expression, with the expected asymptotic behavior $L^* \sim N^{1/d}$ as $N \to \infty$.

 \subsection{Algorithm complexity}

We are now in position to estimate the average-case complexity of the 
range-limited centrality algorithm. For every
root $i$, we sequentially build its $l=1,2,...,L$ shells. When going from shell 
$G_{l-1}(i)$ to building shell $G_{l}(i)$, we consider all the $z_{l-1}$ nodes on 
$G_{l-1}(i)$. For every such node $j$ we add all its links that do not connect to 
already tagged nodes (a tag labels a node that belongs to $G_{l-1}(i)$ or $G_{l-2}(i)$) to 
$G_{l}(i)$, and add the corresponding nodes as well. This requires on the order of
$\langle k \rangle$ operations for every node $j$, hence on the order of 
$\langle k \rangle \langle z_{l-1} \rangle$ operations for creating shell $G_{l}(i)$.
Next is Eq (\ref{sigma}), which involves $\langle e_l\rangle$ steps,
where $e_l$ is the number of edges connecting nodes in shell $G_{l-1}(i)$ to nodes
in shell $G_{l}(i)$. Eq (\ref{bllequal1}) involves $\langle z_l \rangle$ steps.
Eqs (\ref{3a}) and (\ref{3b}) generate a total of $2\sum_{m=1}^l \langle e_m\rangle$
operations. Hence, for a given $l$ there are a total of 
$\langle k \rangle \langle z_{l-1} \rangle + \langle e_l\rangle + \langle z_{l} \rangle +
2\sum_{m=1}^l \langle e_m\rangle$ operations on average. Thus the average 
complexity of the algorithm ${\cal C}$ can be estimated as:
\begin{equation}
{\cal C} \sim  N \sum_{l=1}^L \left( \langle k \rangle \langle z_{l-1} \rangle + 
\langle e_l\rangle + \langle z_{l} \rangle +
2\sum_{m=1}^l \langle e_m\rangle\right) \label{complex1}
\end{equation}
Note that the set of edges in the shells $G_{m-1}(i)$ and $G_{m}(i)$ are all fanning out from
nodes in $G_{m-1}(i)$, and thus we can approximate $\langle e_{m-1}\rangle + 
\langle e_m\rangle$ with $\langle k \rangle \langle z_{m-1} \rangle$. Thus, the estimate 
becomes:
\begin{equation}
{\cal C} \sim N \langle k \rangle \sum_{l=1}^L 
\sum_{m=1}^l \langle z_m\rangle 
= N \langle k \rangle \sum_{l=1}^L (L-l+1) \langle z_{l} \rangle
\label{complex2}
\end{equation}
From (\ref{complex2}) it follows that
\begin{equation}
N \langle k \rangle \sum_{l=1}^L 
\langle z_l\rangle < {\cal C} < L  N \langle k \rangle \sum_{l=1}^L 
\langle z_l\rangle\;. 
\end{equation} 
For {\em fixed} $L$, the complexity grows {\em linearly} with $N$ as $N \to \infty$. 
For  $L=L^*$ we can use (\ref{fit}) to conclude that 
\begin{equation}
{\cal O}(NM)< {\cal C} < {\cal O}(L^* N M) \label{upb}
\end{equation}
where $M = N \langle k \rangle /2$ denotes the total number of edges in the network. 
Recall that the Brandes or Newman algorithm has a complexity of ${\cal O}(NM)$ for
obtaining the traditional betweenness centralities. Specializing the expression 
(\ref{complex2}) to networks with exponentially growing shells one finds the same 
${\cal O}(NM)$ complexity (that is the upper bound ${\cal O}(NM\ln{M})$ 
in (\ref{upb}) is {\em not} realized); for networks with power-law growth shells, however, we find 
${\cal O}(N^{1+1/d}M)$, as in the upper bound of (\ref{upb}). The extra computational
cost is due to the fact that instead of a single 
value, our algorithm 
produces a set of $L$ numbers (the $l$-BCs), providing multiscale information on 
betweenness centrality for all nodes and all edges in the network.

\subsection{Freezing of ranking by range-limited betweenness} \label{sec:freezing}

In Ref \cite{PRLcikkunk}  we have provided numerical evidence that the ranking of 
the nodes (same holds for edges) by their $[L]$-BC values
freezes at relatively small values of $L$. Here we show how this freezing 
phenomenon emerges. Consider two arbitrary nodes $i$ and $j$, 
with degrees $k_i$ and $k_j$.
Using Eq (\ref{nice})
we can write
\begin{equation}
\ln \frac{b_{l}(j)}{b_{l}(i)} = \ln \frac{k_j}{k_i} + \xi_{l}(j) - \xi_{l}(i) = 
 \ln \frac{k_j}{k_i} + \Delta_l\;.  \label{d}
\end{equation} 
 Based on (\ref{xi}):
\begin{equation}
\Delta_l = \xi_{l}(j) - \xi_{l}(i)  = \sum_{n=1}^{l-1} \frac{l+1-n}{l+1} X_n  \label{ixen}
\end{equation}
where $X_n = \epsilon_n(j) - \epsilon_n(i)$. By definition, $\epsilon_n(j)$
is the {\em per node} variation of shell-occupancy from its root-independent
value, for the $n$-th shell centered on root node $j$. 
Expectedly, for larger shells (larger $n$), the size of the shells  
becomes less dependent on the local graph 
structure surrounding the root node, and for this reason this noise term has 
a decaying magnitude $|\epsilon_n(j)|$ with $n$. 
Thus, the $X_n$ can be 
considered as random variables centered around zero, with a magnitude 
that is decaying with increasing $n$. The contributions of the noise terms coming from
larger radius shells in the sum (\ref{ixen}) is decreasing not only because the corresponding
$X_n$-s are decreasing in absolute value, but also because their weight in the sum is 
decreasing (as $1/(l+1)$),
and therefore when moving from $l$ to $l+1$ in (\ref{ixen}) the change (the fluctuation) in 
$\Delta_l$ {\em decreases} for larger $l$.  
This effectively means that the rhs of (\ref{d}) {\em saturates}, and
thus, accordingly, the lhs saturates as well, freezing the ordering of betweenness values. 
If the two nodes have largely different degrees ($\ln k_j/k_i$ is relatively
large), the noise term $\Delta_l$ 
will not be able to change the sign on the rhs of (\ref{d}), even for small $l$ values, and thus,
the ordering between nodes with very different degrees will freeze the fastest, followed by
nodes with degrees that are close to each other. Clearly, the freezing of ordering between
nodes with identical degrees ($k_j = k_i$) will happen last. The probability for the ordering to flip when increasing
the range from $l$ to $l+1$ can be calculated for specific network models, however, it will not
be discussed here.

\section{Range-limited centralities in a large-scale social network} \label{rlcsn}

In this section we illustrate the power of the range-limited approach 
on a real-world social network  inferred from cell-phone call-logs (SocNet).  
We show that computing the $[L]$-BCs up to a relatively small limit length can already
be used to predict the full, diameter-based betweenness centralities 
of individual nodes (and edges), their distribution and the
top list of nodes with highest centralities. 

This social network was constructed from $708$ million anonymized phone-calls between 
$7.2$ million callers generated in a period of $65$ days. 
Restricting ourselves to pairs of individuals between which phone-calls have been observed in both directions in this period as a definition of an edge, we found that the giant 
component of this network has about $5.5$ million nodes and $27$ million edges. 
The $65$ days is long enough to guarantee that individuals with strong social bonds have 
called each other at least once during this interval, and therefore will be linked by an 
edge in our graph.

To test and validate our predictions using the range-limited method, we actually performed the 
computation of the full, diameter-based betweenness centralities of all the nodes in SocNet. 
To deploy the computation, we used a distributed computing utility
called Work Queue, developed in the Cooperative Computing Lab at
Notre Dame. The utility consists of a single management server that
sends tasks out to a collection of heterogeneous workers/processors.
Specifically, our workers consisted of $250$ Sun Grid Engine cores, $300$
Condor cores, and $12$ local workstation cores, for a total of $562$ cores. 
This allowed us to finish thousands of days of computation in the course 
of $5$ days. Each worker received a
request to compute the contribution of shortest paths starting from $50$ vertices 
to the betweenness
centrality of every vertex in the network, summed the $50$ results, 
and sent them back to the management server. Each time the
management server received a contribution, it summed the contribution
with all the others and provided another $50$ vertices for the worker.

We also determined the network diameter from the data using a similar 
distributed computing method, obtaining $D = 26$. 
At first sight this value seems to be at odds with the famous six-degrees of separation phenomenon, 
which implies a much smaller diameter. However, there are two observations 
that one can make here. 1) The social network has a dense core with protruding
branches (``tentacles"), which mathematically speaking, can generate a large diameter. However, 
the experimentally determined six degrees of separation
{\em does not probe} all the branches, it actually relies on the denser core for information flow.
Hence it should be rather similar to the average node-to-node distance, rather than the rigorously defined
network diameter. Indeed, the value of $L^* = 9.35$ that we obtained is rather close to the
six-degrees observation.
2) The social network constructed based on cell-phone communications gives only 
{\em a sample subgraph} of the true social network, where communications happen 
also face-to-face and through land-line phone calls. Hence, one would likely 
measure an even smaller $L^*$ would such data be available.

\subsection{Predicting  betweenness centralities of individual nodes}

In large networks, where measuring the full betweenness centralities 
(i.e., based on all-pair shortest paths)  is too  costly, we can use 
the scaling behavior of range-limited BC values to obtain an estimate 
for the full BC value of a given node. Plotting the $[l]$-BC  values 
measured up to a limit $L$  as function of $l$, we can extrapolate to
ranges beyond $L$. In any finite network the 
$[l]$-BC values will saturate, and thus we expect the appearance of 
finite size effects for large enough $l$, that is in the range $L^*< l \leq D$,
where $L^*$ is the {\em typical radius} of the largest shell and can be 
estimated as described in subsection \ref{L*}. 
\begin{figure}[htbp]  \vspace*{-0.0cm}
\begin{center}
\includegraphics[width=0.34\textwidth]{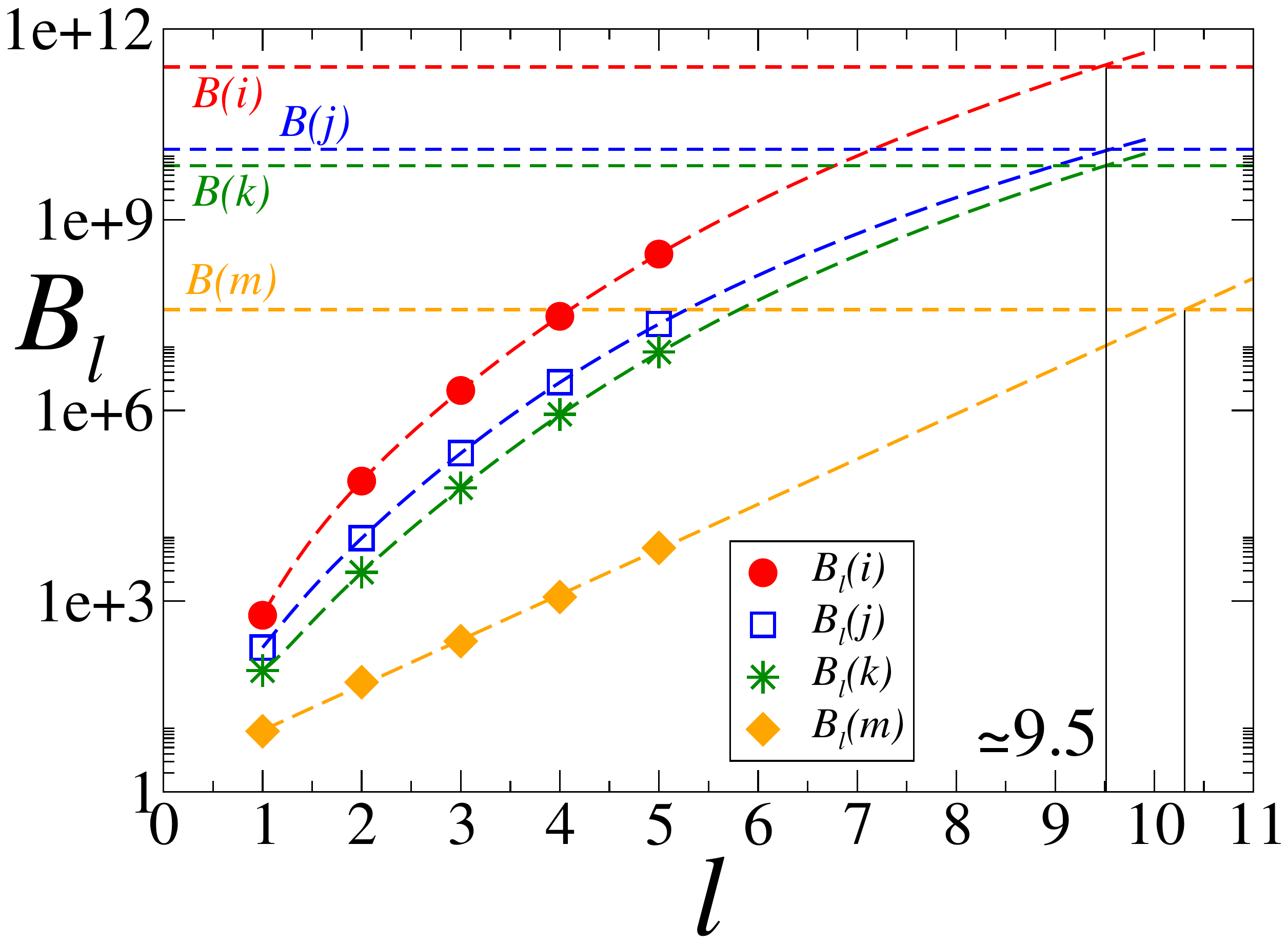} \vspace*{-0.3cm} 
\caption{The $[l]$-BC values, $B_l$, of $4$ individual nodes in 
the SocNet as fucntion of $l$. Range-limited measurements
were made for $l=1,2,3,4,5$, the exact BC value of each 
node is indicated by a horizontal dashed line. Extrapolating
the range-limited values for larger $l$, the real BC value is reached at 
around $9.5$ for nodes $i,j,k$ and $10.3$ for node $m$.
}\label{individB} 
\end{center} \vspace*{-0.2cm} 
 \end{figure}
In Fig.~\ref{individB} we plot the $[l]$-BC values ($B_l(i)$) for $l \leq L=5$ for 
four nodes of SocNet. The four nodes were chosen to have very different $B_l$ values. 
Ranking the nodes by their $[l=5]$-BC values, node $i$ ranked the highest, 
and nodes $j$, $k$ and $m$ ranked $100$, $1000$ and $10000$, respectively. 
The horizontal dashed lines
 represent the full $BC$ values of the nodes obtained from the exact, 
 diameter-length based measurements  (as described above). 
Fitting the five values and extrapolating the range-limited BCs, we can see that  
for nodes $i$, $j$, and $k$, the curves reach their corresponding full BC at 
around $l\simeq 9.5$ agreeing well with the typical length $L^*\simeq 9.35$ 
estimated in subsection \ref{L*}.  
For low ranking nodes (small full BC) finite size effects should appear at lengths  
larger than 
$L^*$,  because they are situated towards the periphery of the graph.
Indeed, one can see from Fig.~\ref{individB} that node $m$ reaches its full 
BC at $l\simeq 10.3$, still fairly close to the estimated $L^*$.

Thus, once we determined $L^*$ as described in \ref{L*}, then by simply 
extrapolating the fitting curve to the $[l]$-BCs of a given node up to  $l = L^*$, 
we obtain an estimate/lower bound for its full betweenness centrality.  

\subsection{Predicting BC distributions}

\begin{figure}[htbp] \vspace*{-0.2cm}
\centerline{\includegraphics[width=0.34\textwidth]{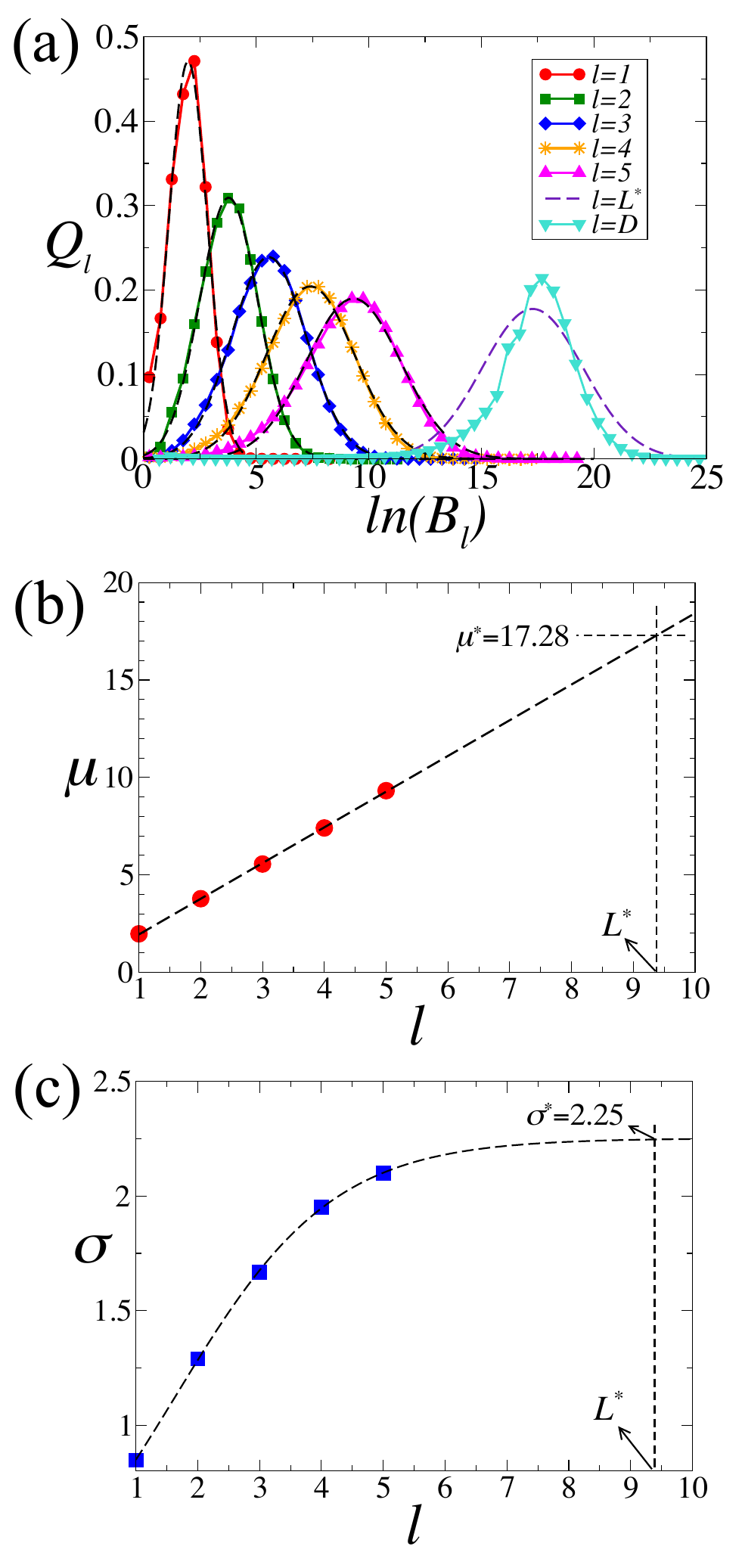}} \vspace*{-0.3cm} 
\caption{a) Distribution $Q_l$ of the $ln(B_l)$ values in the SocNet for $l=1,2,3,4,5,D$, 
where $D=26$ is the diameter, and the predicted distribution
for $L^*$. The distributions can be fitted with a Gaussian.  b)The average $\mu$ and 
c) standard deviation $\sigma$ as function of $l$. Extrapolating to 
$L^*=9.35$ we obtain $\mu^*=17.28$ 
and $\sigma^*=2.25$.}\label{distrib} 
\vspace*{-0.3cm} 
 \end{figure}

In SocNet the  $B_l$ values have a lognormal distribution \cite{PRLcikkunk}, 
thus $Q_l(ln(B_l))$  can be well fitted 
by a Gaussian (Fig.~\ref{distrib}a). The parameters
of the distribution also show a scaling behavior, and extrapolating up to $L^*=9.35$ 
we obtain $\mu^*=17.28$ for the average (Fig.~\ref{distrib}b)
and $\sigma^*=2.25$ (Fig.~\ref{distrib}c) for the standard deviation of the Gaussian. 
This predicted distribution 
is shown as a dashed line on Fig.~\ref{distrib}a. Comparing it with the distribution of the 
full BC values ($l=D$) we can see that while the averages
agree, the width of the distribution is, however, smaller than the predicted value. 
This is caused by the fact, that the $[l]$-BCs  do not saturate at the same $l$ value: for 
low centrality nodes saturation occurs at larger $l$, as also shown in Fig.~\ref{individB}. 

\subsection{Predicting BC ranking}

Efficiently identifying high betweenness centrality nodes and edges is rather important in many
applications, as these nodes and edges both handle large-amounts of traffic (thus 
they can be bottlenecks or congestion hotspots),  and  form high-vulnerability 
subsets  (their removal may lead to major failures).  Fortunately, due to the freezing
phenomenon described in subsection \ref{sec:freezing}, one does not need to
compute accurately the full BC-s in order to identify the top ranking nodes and
edges. At already modest $l$ values we obtain top-lists that have a strong overlap with the
ultimate,  $[l=D]$-BC top-list. Here we illustrate this for the case of SocNet. 
\begin{table*}[htbp]
%\centering
 % \begin{minipage}{10.0cm}
  \begin{tabular}{|c|c|c|c|c|c|c|}
    \hline    
    Vertex          & $B_1$   & $B_2$   & $B_3$  & $B_4$     & $B_5$ & $B_D$     \\
    \hline 
    1    &   \, 600 \,   & \,$7.76\times 10^4$\, & \,$2.06\times10^6$\, & \,$3.01\times 10^7$\, & \,$2.87\times 10^8$\,& \,$1.26\times 10^{11}$\,\\
    \hline
    2    &   715    & $9.71\times 10^4$ & $2.22\times10^6$& $3.05\times10^7$ & $2.85\times10^8$& $1.25\times10^{11}$\\
    \hline
    3    &      458  &  $5.04\times 10^4$ & $1.26\times10^6$& $1.87\times10^7$ & $1.86\times10^8$& $9.82\times10^{10}$ \\
    \hline
    4    &      377  &  $3.11\times 10^4$ & $8.56\times10^5$& $1.31\times10^7$ & $1.29\times10^8$& $5.82\times10^{10}$  \\
    \hline
    5    &      337  & $2.29\times 10^4$ & $5.04\times10^5$& $7.23\times10^6$ & $7.55\times10^7$& $5.34\times10^{10}$ \\
    \hline
      6    &      285  & $1.93\times 10^4$ & $5.03\times10^5$& $7.56\times10^6$ & $7.85\times10^7$& $5.07\times10^{10}$ \\
    \hline
      7    &      488  & $2.82\times 10^4$ & $5.84\times10^5$& $7.94\times10^6$ & $7.96\times10^7$& $4.89\times10^{10}$ \\
    \hline
      8    &     299   & $2.56\times 10^4$ & $6.91\times10^5$& $1.09\times10^7$ & $1.10\times10^8$& $4.88\times10^{10}$ \\
    \hline
      9    &      244  & $1.47\times 10^4$ & $3.44\times10^5$& $4.87\times10^6$ & $4.83\times10^7$& $4.87\times10^{10}$ \\
    \hline
      10    &      239  & $1.64\times 10^4$ & $4.57\times10^5$& $7.48\times10^6$ & $8.06\times10^7$& $4.81\times10^{10}$\\
    \hline
  \end{tabular}
  \caption{$B_l$ values of the top $10$ nodes in the $[D]$-BC top-list for SocNet, 
  for $l=1,2,3,4,5,D$, where  $D=26$ is the diameter. \label{ranking}}
\end{table*}
 % \end{minipage}
 % \begin{minipage}{6.5cm}
 \begin{table}[htbp]
  \begin{tabular}{|c|c|}
    \hline    
           \,  Top $x$ nodes \, & \,Overlap ($\%$)\,    \\
    \hline
   1 & 100 \\
    \hline
 2 & 100   \\
    \hline
3 & 100  \\
    \hline
4 & 100  \\
    \hline
 10 & 90    \\
    \hline
    50 & 72    \\
    \hline
    100 & 75    \\
    \hline
    500 & 70.2    \\
    \hline
    1000 & 67.1    \\
    \hline
  \end{tabular}
  \caption{Overlap between the lists of the top $r$ nodes with highest $[5]$-BC and  
  with the highest $[D]$-BC values. \label{rankcorr} }
%  \end{minipage}
\end{table}
Table \ref{ranking} lists the $[l]$-BC (for $l=1,2,3,4,5$ and  $l=D=26$) 
of the top $10$ nodes from the $[D]$-BC list in SocNet.
The overlap between the  top lists at consecutive $l$ values increases with $l$.
Given two lists, we define the overlap between their first (top-ranking) $r$ elements by the
percentage of common elements in both $r$-element lists.  
Table \ref{rankcorr}  shows the overlap between the top list based on  
$[5]$-BC and the one based on
the ultimate $[D]$-BC values. At $l=5$ the top $4$ nodes are already exactly in 
the same order as in the $[D]$-BC list, the overlap is $90\%$ between the lists of the 
top $10$ nodes, and even for the top 
$100$ node lists we have an overlap of $75\%$.

\section{Range-limited centrality in weighted graphs} \label{rlcwn}

In unweighted graphs the length of the shortest path between 
two nodes is defined as the number of edges included in the shortest 
path. In weighted networks each edge has a weight or  ``length'': $w_{ij}$. 
Depending on the nature of the network this length can be an actual 
physical distance  (e.g., in road networks), or a cost or a resistance value. 
We define the ``shortest'' (or lowest-weight) path between nodes 
$i$ and $j$ as the network path along which the sum of the weights of the edges 
included is minimal. We will call this sum as the ``shortest distance'' $d(i,j)$  from node $i$ to node 
$j$ (note that we allow for directed links, which implies that $d(i,j)$ is not necessarily the same
as $d(j,i)$). 
 
In order to define a range-limited quantity, let $b_l(j)$ denote the (fixed) $l$-BC of node $j$ 
from all-pair 
shortest directed paths of length $W_{l-1}< d \leq W_l$, where $W_1<W_2< \dots <W_L$ 
are a series of predefined weight values or ``distances''. The simplest way to define these 
$W_l$ distances is to take them uniformly $W_l=l \Delta w$, however depending on the application 
these may be redefined in any suitable way.  $B_L$ will again denote the cumulative 
$L$-betweenness, which represents centralities from paths not longer than $W_L$. Note that
we are still counting paths when computing centralities, that is $\sigma_{mn}(i)$ still means
the number of shortest paths from $m$ to $n$ passing through $i$, except for the meaning of
``shortest", which is now generalized to lowest-cost. 
 
\begin{figure}[htbp] \begin{center}
\includegraphics[width=0.47\textwidth]{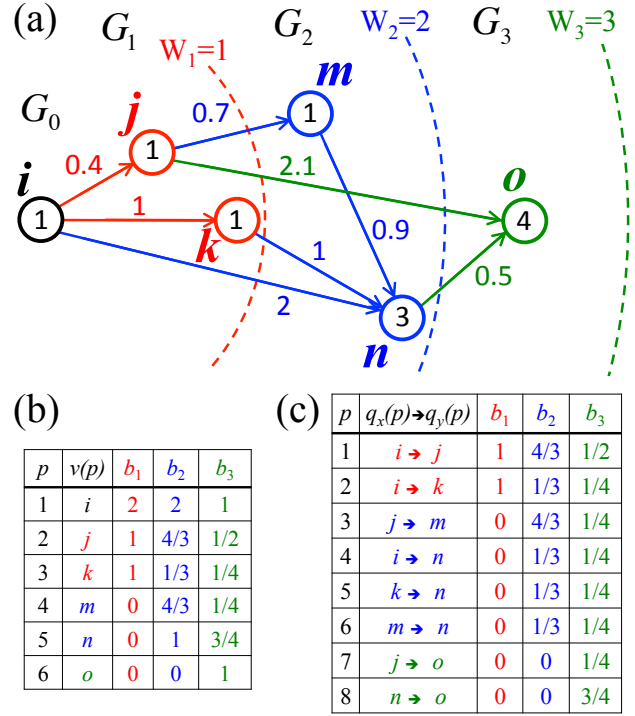}
\caption{a) Shells of the $\bm{C}_3$ subgraph of node $i$ (black) are colored 
red, blue, green. Distances defining the shells are: $W_1=1$, $W_2=2$, $W_3=3$. 
The weight or length is shown next to each edge.
Given a node $j$, the number inside its circle 
  is the total number of shortest paths coming from the root $i$: $\sigma_{ij}$.
b) The list of nodes $v(p)$ and c) list of edges $q_x(p)\rightarrow q_y(p)$ are shown 
together with their $1$-, $2$-, and $3$- betweenness values.}\label{figweighted} 
\vspace*{-0.5cm} \end{center}
 \end{figure}

The algorithm is similar to the one presented above for unweighted networks. 
We again build the subgraph of a node $i$, but now a shell  $G_l(i)$ will contain 
all the nodes $k$ at shortest path  distance $W_{l-1}< d(i,k) \leq W_l$ from the root node $i$.  
An edge $j\rightarrow k$ is considered to be part of the layer in which node $k$ is included.
In unweighted graphs a connection $j\rightarrow k$ can be part of the 
subgraph  only if the two nodes are in two consecutive layers: if $j\in G_r(i)$ 
then $k\in G_{r+1}(i)$. In weighted networks the situation is
different (Fig.~\ref{figweighted}a)). In principle we may have edges connecting 
nodes which are not in two consecutive layers, but possibly further away from each other 
( the links $i\rightarrow n$, $j\rightarrow o$ in Fig.~\ref{figweighted}a)), or  
even in the same layer (the link $m\rightarrow n$ in the same figure). 

When building the subgraph using breadth-first search, we need to save the exact order in which
the nodes and edges are discovered and included in the subgraph (Fig.~\ref{figweighted}b,c). 
Let us denote with $v(p)$ the index of the node which is included at position $p$ in this node's 
list (Fig.~\ref{figweighted}b). This means that 
the following conditions hold: $d(i,v(1))\leq d(i,v(2))\leq d(i,v(3)) \leq \dots$. 
Similarly we have a list of edges, where $q_x(p)\rightarrow q_y(p)$ is the edge in position 
$p$ of the list, and $q_x$, $q_y$ denote the indexes of the two nodes connected 
by the edge (Fig.~\ref{figweighted}c). This implies the conditions: $d(i,q_y(1))\leq d(i,q_y(2))\leq d(i,q_y(3)) \leq \dots$ (note that every  
 edge $q_x(p)\rightarrow q_y(p)$  is  included in the edge-list when node $q_y(p)$ is discovered).
Again, we calculate  $b_l^r(i|k)$ for a node $k$, and  $b_l^r(i|j,k)$ for an edge $j\rightarrow k$. 
As defined above, these values take into account only the shortest paths starting from node $i$, 
and $r$ denotes the shell containing the corresponding node or edge.
One uses the same initial conditions $\sigma_{ii}=1$, and $\sigma_{ik}=0$ for all $k\neq i$, as before.

The algorithm has the following main steps. For every $l=1,\dots,L$:

1) We build the next layer $G_l(i)$ using breadth first search.
During this search we build the list of indexes $v$, $q_x$, $q_y$ 
 as defined above. We denote the total number of nodes 
 included in the list (from all shells $G_1(i)$ 
 up to $G_l(i)$) as $N_l$ and the number of edges included as $M_l$.
During this breadth-first search we also calculate the $\sigma_{ik}$ 
of the discovered nodes. Every time a new edge
$j\rightarrow k$ is added to the list we update $\sigma_{ik}$ by 
adding to it $\sigma_{ij}$ (using algorithmic notation, $\sigma_{ik}:=\sigma_{ik}+\sigma_{ij}$). 
Recall that $\sigma_{ik}$ denotes
 the total {\em number} of shortest paths from $i$ to $k$. If  the  
 edge $j\rightarrow k$ is included in the subgraph 
(meaning that it is part of a shortest path) the number of shortest 
 paths ending in $j$ has to be added to the 
 number of shortest paths ending in $k$. 

2) The $l$-betweenness of all nodes included in the new layer is set to $b_l^l(i|k)=1$, similarly to Eq. (\ref{bllequal1}).

3) Going backwards through the list of edges we calculate the fixed-$l$ BC of all nodes and edges. For $p=M_l,\dots,1$,
 we perform the following recursions:
 
 a) for the edge $q_x(p)\rightarrow q_y(p)$:
\begin{equation}
b_l^r(i|q_x(p),q_y(p))=b_l^r(i|q_y(p))\frac{\sigma_{iq_x(p)}}{\sigma_{iq_y(p)}}
\end{equation}
b) {\it immediately after}  the BC of an edge is calculated, the  betweenness of 
node $q_x(p)$ must also be updated. We have to 
add to its previous value the $l$-BC of the edge $q_x(p)\rightarrow q_y(p)$:
\begin{equation}
b_l^r(i|q_x(p))=b_l^r(i|q_x(p))+b_l^r(i|q_x(p),q_y(p))
\end{equation}

4) We return to step 1) until the last shell $G_L(i)$ is reached.

As we have seen, the algorithm and the recursions are very similar to the one presented for unweighted graphs. The
crucial difference is that the exact order of the discovered nodes and edges has to be saved, 
because the BC values of edges and nodes in a shell cannot be updated in an arbitrary order.
As an example, Fig.~\ref{figweighted}  shows a small subgraph and the list of nodes and 
edges together with their $1$-, $2$- and $3$-betweenness values. 

\begin{figure}[htbp] \begin{center}
\includegraphics[width=0.48\textwidth]{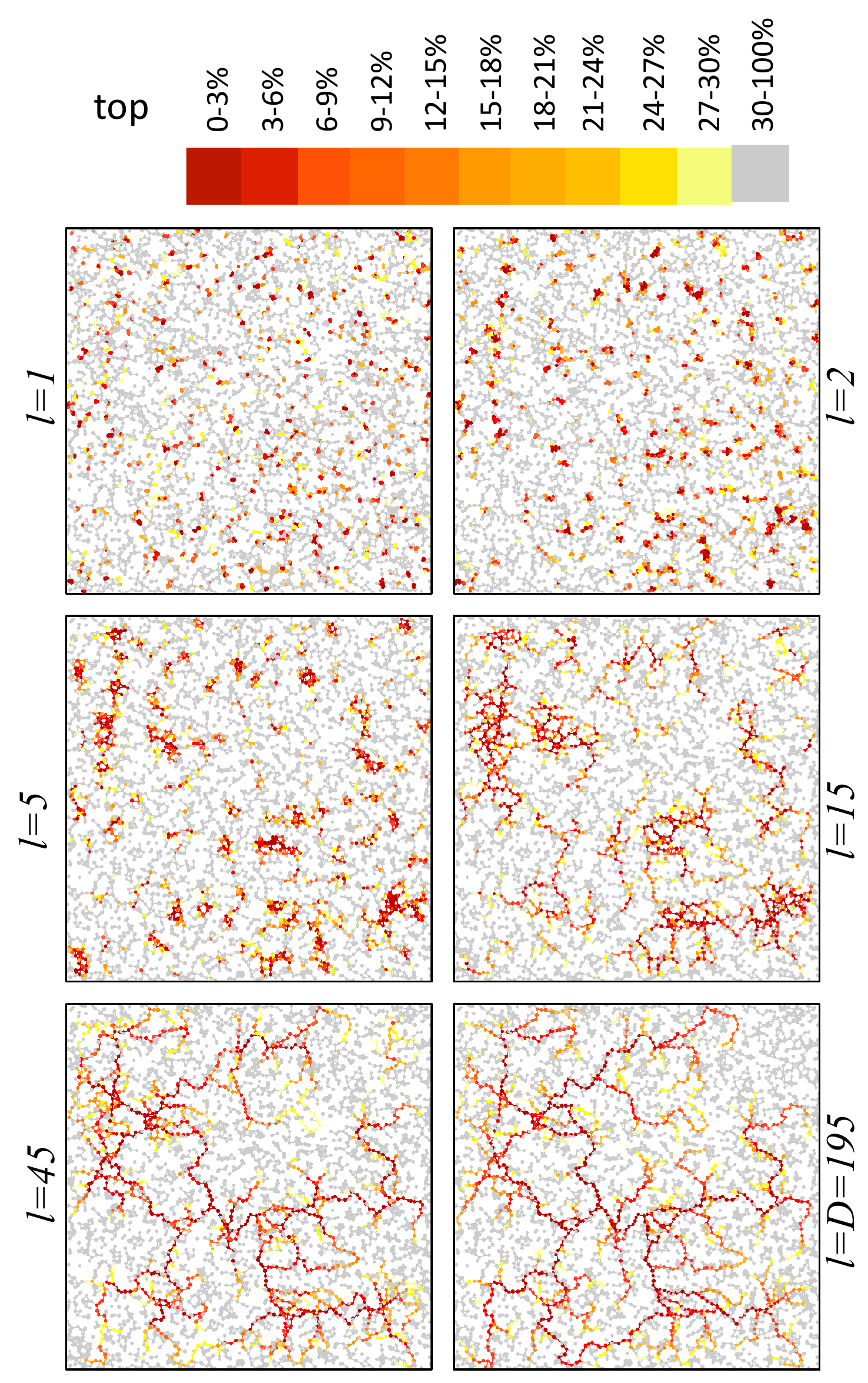}
\caption{The vulnerability backbone VB of a random geometric graph in the unit square
with $N=5000$, $ \langle k \rangle=5$ and $D=195$.   The top $30\%$ of nodes are 
colored from red to yellow according to their $[l]$-BC ranking (see color bar). 
The VB based on the $[l]$-BC is shown for different values: $l=1,2,5,15,45,195$. }
\label{backbone}  \vspace*{-0.5cm} \end{center}
 \end{figure}

\section{Vulnerability backbone} \label{vbb}

An important problem in network research is identifying the most vulnerable parts of a network. 
Here we define the vulnerability backbone (VB) of a graph as the smallest fraction of the 
highest betweenness nodes {\em forming a percolating cluster} through the network. Removing
simultaneously all elements of this backbone will efficiently 
shatter the network into many disconnected  pieces. Although the shattering performance
can be improved by sequentially removing and {\em recomputing} the top-ranking nodes 
\cite{holme}, here we focus only on the simultaneous removal of the
one-time computed VB of a graph, the generalization being straightforward.

Next we illustrate that range-limited BCs can be used to efficiently detect this backbone by performing calculations up to a length much smaller than the diameter. This is of course 
expected in  networks that have a small diameter ($D={\cal O}(\ln N)$ or smaller), however, it 
is less obvious for networks with large diameter ($D={\cal O}(N^\alpha)$, $\alpha > 0$). 
For this reason, in the following we consider random geometric (RG) graphs \cite{penrose,RG}
in the plane. The graphs are obtained by sprinkling at random $N$ points into the unit 
square and connecting all pairs of points that are found within a given distance 
$R$ of each other. We will use the average degree $\langle k\rangle = N\pi R^2$ \cite{RG}  
instead of $R$ to parametrize the graphs. In Fig. \ref{backbone} we present measurements on
a random  geometric graph with $N=5000$ nodes,
average degree $\langle k \rangle=5$. The hop-count diameter of this graph is $D=195$. 
The weights of connections are considered to be the physical (Euclidean) distances. Clearly, since the links
of the graph are built based on a rule involving the Euclidean distances, the weight structure
and the topology of the graph should be tightly correlated. Thus, we do expect 
strong correlations between the $[l]$-BC values measured both from the unweighted and 
the weighted graph.
The weight ranges $W_l$ defining the layers during the algorithm 
were chosen as $W_l=0.00725 l$, $l=1,\ldots, D$, so that $W_D= 0.00725 D =1.413$ is 
close to the diagonal length of the unit square $\sqrt{2}$. 
The nodes and connections are colored according to their $[l]-BC$ ranking 
for different $l$ values
(see the color bar in Fig \ref{backbone}).  
The backbone is already clearly formed at $l=45$. 
\begin{figure}[htbp] \begin{center}
\includegraphics[width=0.48\textwidth]{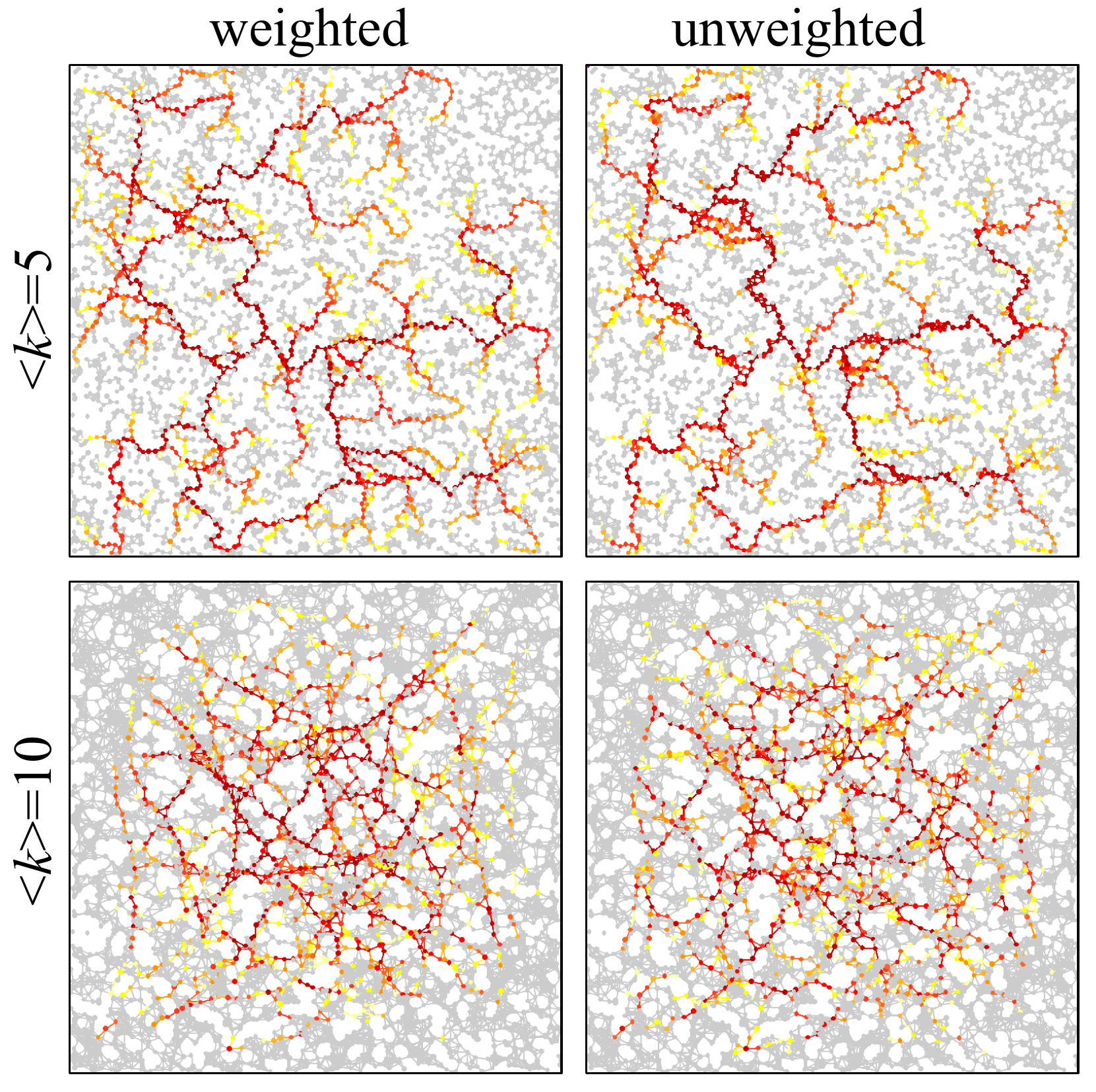}
\caption{Vulnerability backbones based on full BC rankings 
in two random geometric graphs with $N=5000$ nodes, 
and average degrees $\langle k \rangle=5$ and $\langle k \rangle=10$,
respectively.  The rankings were calculated both on the unweighted graph 
(left column) and weighted one (right column). } \label{backbonecomp} 
\vspace*{-0.5cm} \end{center}
 \end{figure}
Fig. \ref{backbonecomp} compares the VBs  of the
graphs obtained with and without considering the connection weights (distances). Two RGs 
with densities $\langle k \rangle=5$ and $\langle k \rangle=10$ are presented. In the 
case of the denser graph the backbone is concentrated towards the center of unit
square,
as periphery effects in this case are stronger (we do not use periodic boundary conditions).
\begin{figure}[htbp] \begin{center}\vspace*{-0.3cm} 
\includegraphics[width=0.48\textwidth]{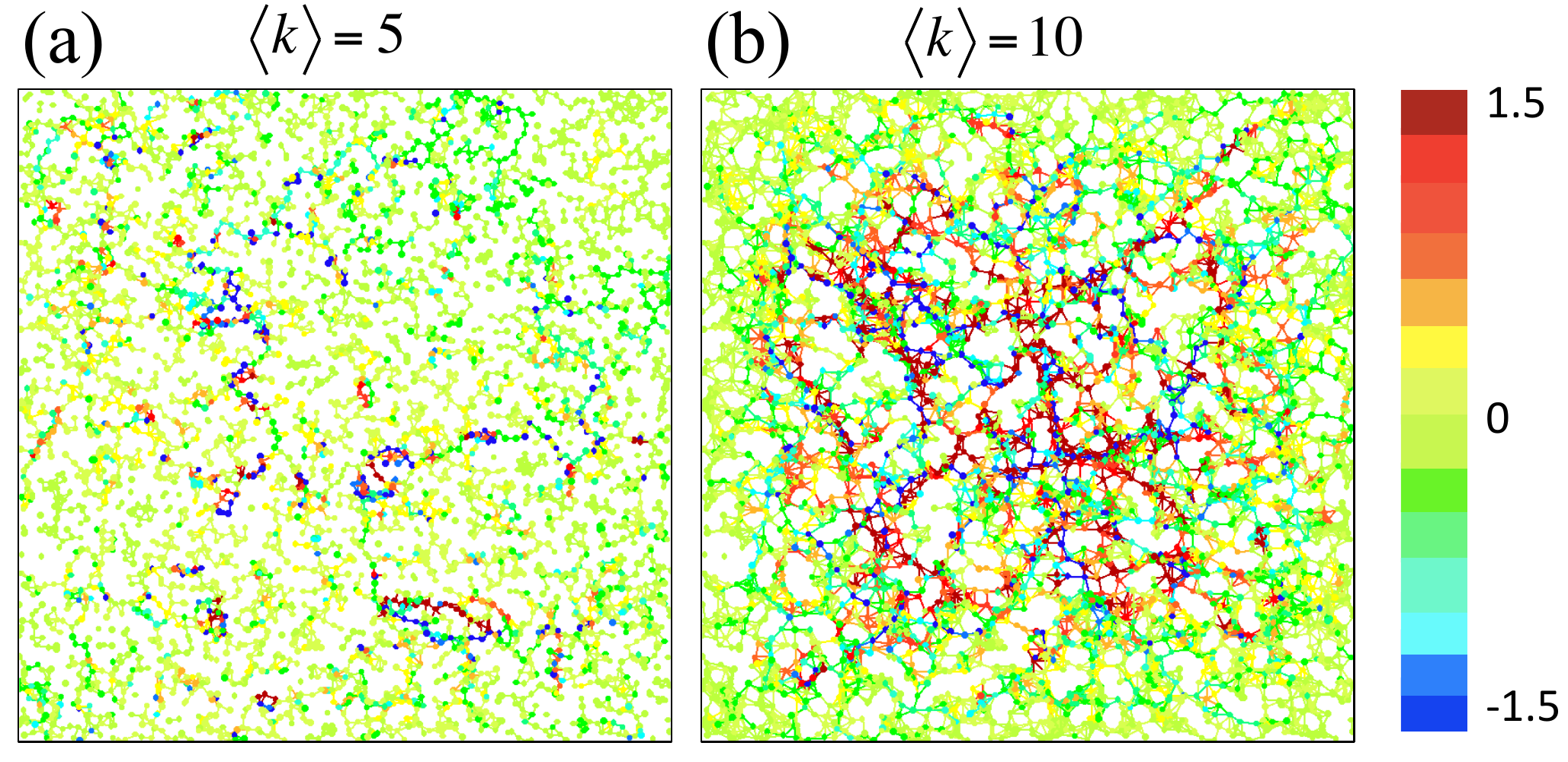}\vspace*{-0.2cm} 
\caption{Comparison between the rankings obtained with and without considering 
the weights of connections for the two RG graphs
in Fig. \ref{backbonecomp}. Colors indicate the $ln(r_{nw}/r_{w})$ values, where 
$r_{nw}$ is the rank of a node 
obtained using the non-weighted algorithm and $r_w$ is obtained with the weighted 
graph (see the color bar). In denser graphs the differences become more significant.} \label{rankdiff} 
\vspace*{-0.7cm} \end{center}
 \end{figure}
Although qualitatively the two VBs are similar, the VB is sharper and clearer in 
the weighted case. There can be actually significant differences between the two 
backbones, in spite the fact that
one would expect a strong overlap.  In Fig. \ref{rankdiff} we show these 
differences by coloring the nodes of the two 
graphs from Fig \ref{backbonecomp} according to the 
$ln(r_{nw}/r_{w})$ values, where  $r_{nw}$ is the rank of a node 
obtained using the non-weighted algorithm and $r_w$ is obtained using the 
weighted graph. The nodes are
colored from blue to red, blue corresponding to the case when the unweighted algorithm strongly 
underestimates the weighted ranking of a node and red is used when it overestimates it. 
Although it is of no surprise that  weighted and unweighted backbones differ in networks
where the graph topology and the weights are weakly correlated, the fact that there are considerable differences also for the strongly correlated case of random geometric graphs
(the blue and red colored parts in the right panel of Fig  \ref{rankdiff}) is rather unexpected, underlining 
the importance of using weigh-based centrality measures in weighted networks.

\section{Conclusions}

In this paper we have introduced a systematic approach to network centrality measures decomposed by graph distances for both unweighted and weighted
directed networks. There are several advantages to such range-based decompositions.
First, they provide much finer grained information on the positioning importance of a 
node (or edge) with respect to the network, than the traditional (diameter-based) centrality measures. Traditional centrality values are dominated by the 
large number of long-distance network paths, even though most of these paths might not
actually be used frequently by the transport processes occurring on the network. 
Due to the fast growth of the number of paths with distance in large complex networks, one expects that the distribution of the centrality measures (which incorporate these paths) to obey scaling laws as
the range is increased.
We have shown both numerically and via analytic arguments (identifying the scaling form) 
that this is indeed the case, for unweighted networks;  for the same reasons, however, 
we expect the existence of scaling laws for weighted networks as well. 
We have shown that these scaling laws can be used to predict
or estimate efficiently several quantities of interest, that are otherwise costly to compute 
on large networks. In particular, the largest typical node-to-node distance $L^*$, the traditional
individual node and edge centralities (diameter range) and the ranking of nodes and edges by their
centrality values. The latter is made possible by the existence of the phenomenon of fast
freezing of the rank ordering by distance, which we demonstrated both numerically and via
analytic arguments. 
We have also introduced efficient algorithms for range-limited centrality measures 
for both unweighted and weighted networks. Although they have been presented for
betweenness centrality, they can be modified to obtain all the other centrality measure
variants.  Finally, we presented an application of these concepts in identifying the vulnerability 
backbone of a network, and have shown that it can be identified efficiently using range-limited
betweenness centralities. We have also illustrated the importance of taking into 
account link-weights \cite{PNAS_SBV09} when computing centralities, even in 
networks where graph topology and weights are strongly correlated.

\section*{Acknowledgments}
\noindent This work was supported in part by  NSF BCS-0826958, 
HDTRA 1-09-1-0039 and by the Army Research Laboratory
under Cooperative Agreement Number W911NF-09-2-0053 and MER in part by  
PN-II-RU-TE-2011-3-0121.
The views and
conclusions contained in this document are those of the authors and should
not be interpreted as representing the official policies, either expressed
or implied, of the Army Research Laboratory or the U.S. Government.  The
U.S. Government is authorized to reproduce and distribute reprints for
Government purposes notwithstanding any copyright notation here on.

%\bibliography{betwPREuj.bib}

\begin{thebibliography}{75}
\expandafter\ifx\csname natexlab\endcsname\relax\def\natexlab#1{#1}\fi
\expandafter\ifx\csname bibnamefont\endcsname\relax
  \def\bibnamefont#1{#1}\fi
\expandafter\ifx\csname bibfnamefont\endcsname\relax
  \def\bibfnamefont#1{#1}\fi
\expandafter\ifx\csname citenamefont\endcsname\relax
  \def\citenamefont#1{#1}\fi
\expandafter\ifx\csname url\endcsname\relax
  \def\url#1{\texttt{#1}}\fi
\expandafter\ifx\csname urlprefix\endcsname\relax\def\urlprefix{URL }\fi
\providecommand{\bibinfo}[2]{#2}
\providecommand{\eprint}[2][]{\url{#2}}

\bibitem[{\citenamefont{Cohen and Havlin}(2010)}]{CohenHavlinBook}
\bibinfo{author}{\bibfnamefont{R.}~\bibnamefont{Cohen}} \bibnamefont{and}
  \bibinfo{author}{\bibfnamefont{S.}~\bibnamefont{Havlin}},
  \emph{\bibinfo{title}{Complex Networks: Structure, Robustness and Function}}
  (\bibinfo{publisher}{Cambridge Univ. Press}, \bibinfo{year}{2010}).

\bibitem[{\citenamefont{Newman}(2010)}]{NewmanBook}
\bibinfo{author}{\bibfnamefont{M.}~\bibnamefont{Newman}},
  \emph{\bibinfo{title}{Networks: An Introduction}} (\bibinfo{publisher}{Oxford
  University Press, USA}, \bibinfo{year}{2010}).

\bibitem[{\citenamefont{Barrat and Vespignani}(2008)}]{BarratVespBook}
\bibinfo{author}{\bibfnamefont{A.}~\bibnamefont{Barrat}} \bibnamefont{and}
  \bibinfo{author}{\bibfnamefont{A.}~\bibnamefont{Vespignani}},
  \emph{\bibinfo{title}{Dynamical Processes on Complex Networks}}
  (\bibinfo{publisher}{Cambridge Univ. Press}, \bibinfo{year}{2008}).

\bibitem[{\citenamefont{Boccaletti et~al.}(2006)\citenamefont{Boccaletti,
  Latora, Moreno, Chavez, and Hwang}}]{BoccaEtAlRev06}
\bibinfo{author}{\bibfnamefont{S.}~\bibnamefont{Boccaletti}},
  \bibinfo{author}{\bibfnamefont{V.}~\bibnamefont{Latora}},
  \bibinfo{author}{\bibfnamefont{Y.}~\bibnamefont{Moreno}},
  \bibinfo{author}{\bibfnamefont{M.}~\bibnamefont{Chavez}}, \bibnamefont{and}
  \bibinfo{author}{\bibfnamefont{D.-U.} \bibnamefont{Hwang}},
  \bibinfo{journal}{Phys. Rep.} \textbf{\bibinfo{volume}{424}},
  \bibinfo{pages}{175} (\bibinfo{year}{2006}).

\bibitem[{\citenamefont{Ben-Naim et~al.}(2004)\citenamefont{Ben-Naim,
  Frauenfelder, and Toroczkai}}]{BenFrauTor04}
\bibinfo{editor}{\bibfnamefont{E.}~\bibnamefont{Ben-Naim}},
  \bibinfo{editor}{\bibfnamefont{F.}~\bibnamefont{Frauenfelder}},
  \bibnamefont{and}
  \bibinfo{editor}{\bibfnamefont{Z.}~\bibnamefont{Toroczkai}}, eds.,
  \emph{\bibinfo{title}{Complex Networks}}, Lecture Notes in Physics
  (\bibinfo{publisher}{Springer-Verlag, Berlin}, \bibinfo{year}{2004}).

\bibitem[{\citenamefont{Wasserman and Faust}(1994)}]{wasserman}
\bibinfo{author}{\bibfnamefont{S.}~\bibnamefont{Wasserman}} \bibnamefont{and}
  \bibinfo{author}{\bibfnamefont{K.}~\bibnamefont{Faust}},
  \emph{\bibinfo{title}{Social Network Analysis: methods and applications}}
  (\bibinfo{publisher}{Cambridge Univ. Press}, \bibinfo{year}{1994}).

\bibitem[{\citenamefont{Scott}(1991)}]{scott}
\bibinfo{author}{\bibfnamefont{J.}~\bibnamefont{Scott}},
  \emph{\bibinfo{title}{Social Network Analysis: A Handbook}}
  (\bibinfo{publisher}{Sage Publications}, \bibinfo{year}{1991}).

\bibitem[{\citenamefont{Sabidussi}(1966)}]{sabidussi}
\bibinfo{author}{\bibfnamefont{G.}~\bibnamefont{Sabidussi}},
  \bibinfo{journal}{Psychometrika} \textbf{\bibinfo{volume}{31}},
  \bibinfo{pages}{581} (\bibinfo{year}{1966}).

\bibitem[{\citenamefont{Friedkin}(1991)}]{friedkin}
\bibinfo{author}{\bibfnamefont{N.~E.} \bibnamefont{Friedkin}},
  \bibinfo{journal}{Amer. J. of Soc.} \textbf{\bibinfo{volume}{96}},
  \bibinfo{pages}{1478} (\bibinfo{year}{1991}).

\bibitem[{\citenamefont{Borgatti and Everett}(2006)}]{SocNet_BE06}
\bibinfo{author}{\bibfnamefont{S.~P.} \bibnamefont{Borgatti}} \bibnamefont{and}
  \bibinfo{author}{\bibfnamefont{M.~G.} \bibnamefont{Everett}},
  \bibinfo{journal}{Social Networks} \textbf{\bibinfo{volume}{28}},
  \bibinfo{pages}{466} (\bibinfo{year}{2006}).

\bibitem[{\citenamefont{Anthonisse}(1971)}]{anthonisse}
\bibinfo{author}{\bibfnamefont{J.~M.} \bibnamefont{Anthonisse}},
  \bibinfo{journal}{Tech. Rep. BN 9/71, Stichting Math. Centr., Amsterdam}
  (\bibinfo{year}{1971}).

\bibitem[{\citenamefont{Freeman}(1977)}]{freeman77}
\bibinfo{author}{\bibfnamefont{L.~C.} \bibnamefont{Freeman}},
  \bibinfo{journal}{Sociometry} \textbf{\bibinfo{volume}{40}},
  \bibinfo{pages}{35} (\bibinfo{year}{1977}).

\bibitem[{\citenamefont{Freeman}(1979)}]{SocNet_F79}
\bibinfo{author}{\bibfnamefont{L.~C.} \bibnamefont{Freeman}},
  \bibinfo{journal}{Soc. Netw.} \textbf{\bibinfo{volume}{1}},
  \bibinfo{pages}{215} (\bibinfo{year}{1979}).

\bibitem[{\citenamefont{Brandes}(2008)}]{SocNet_B08}
\bibinfo{author}{\bibfnamefont{U.}~\bibnamefont{Brandes}},
  \bibinfo{journal}{Soc. Netw.} \textbf{\bibinfo{volume}{30}},
  \bibinfo{pages}{136} (\bibinfo{year}{2008}).

\bibitem[{\citenamefont{White and Borgatti}(1994)}]{WhiteBorgatti94}
\bibinfo{author}{\bibfnamefont{D.}~\bibnamefont{White}} \bibnamefont{and}
  \bibinfo{author}{\bibfnamefont{S.}~\bibnamefont{Borgatti}},
  \bibinfo{journal}{Soc. Netw.} \textbf{\bibinfo{volume}{16}},
  \bibinfo{pages}{335} (\bibinfo{year}{1994}).

\bibitem[{\citenamefont{Borgatti}(2005)}]{SocNet_B05}
\bibinfo{author}{\bibfnamefont{S.}~\bibnamefont{Borgatti}},
  \bibinfo{journal}{Soc. Networks} \textbf{\bibinfo{volume}{27}},
  \bibinfo{pages}{55 } (\bibinfo{year}{2005}).

\bibitem[{\citenamefont{Sreenivasan et~al.}(2007)\citenamefont{Sreenivasan,
  Cohen, Lopez, Toroczkai, and Stanley}}]{PRE_SCTS07}
\bibinfo{author}{\bibfnamefont{S.}~\bibnamefont{Sreenivasan}},
  \bibinfo{author}{\bibfnamefont{R.}~\bibnamefont{Cohen}},
  \bibinfo{author}{\bibfnamefont{E.}~\bibnamefont{Lopez}},
  \bibinfo{author}{\bibfnamefont{Z.}~\bibnamefont{Toroczkai}},
  \bibnamefont{and} \bibinfo{author}{\bibfnamefont{H.~E.}
  \bibnamefont{Stanley}}, \bibinfo{journal}{Phys.Rev.E}
  \textbf{\bibinfo{volume}{75}}, \bibinfo{pages}{036105}
  (\bibinfo{year}{2007}).

\bibitem[{\citenamefont{Dolev et~al.}(2010)\citenamefont{Dolev, Elovici, and
  Puzis}}]{JACM_DEP10}
\bibinfo{author}{\bibfnamefont{S.}~\bibnamefont{Dolev}},
  \bibinfo{author}{\bibfnamefont{Y.}~\bibnamefont{Elovici}}, \bibnamefont{and}
  \bibinfo{author}{\bibfnamefont{R.}~\bibnamefont{Puzis}}, \bibinfo{journal}{J.
  ACM} \textbf{\bibinfo{volume}{57}} (\bibinfo{year}{2010}).

\bibitem[{\citenamefont{Bollob\'as}(1991)}]{BollobasBook}
\bibinfo{author}{\bibfnamefont{B.}~\bibnamefont{Bollob\'as}},
  \emph{\bibinfo{title}{Modern Graph Theory}}, Graduate Texts in Mathematics
  (\bibinfo{publisher}{Springer-Verlag, New York, Berlin Heidelberg},
  \bibinfo{year}{1991}).

\bibitem[{\citenamefont{Shimbel}(1953)}]{shimbel}
\bibinfo{author}{\bibfnamefont{A.}~\bibnamefont{Shimbel}},
  \bibinfo{journal}{Bulletin of Math. Biophys.} \textbf{\bibinfo{volume}{15}},
  \bibinfo{pages}{501} (\bibinfo{year}{1953}).

\bibitem[{\citenamefont{Perer and Shneiderman}(2006)}]{Perer2006}
\bibinfo{author}{\bibfnamefont{A.}~\bibnamefont{Perer}} \bibnamefont{and}
  \bibinfo{author}{\bibfnamefont{B.}~\bibnamefont{Shneiderman}},
  \bibinfo{journal}{IEEE Trans. on Visualization and Computer Graphics}
  \textbf{\bibinfo{volume}{12}}, \bibinfo{pages}{693} (\bibinfo{year}{2006}).

\bibitem[{\citenamefont{Lammer et~al.}(2006)\citenamefont{Lammer, Gehlsen, and
  Helbing}}]{Lammer2006}
\bibinfo{author}{\bibfnamefont{S.}~\bibnamefont{Lammer}},
  \bibinfo{author}{\bibfnamefont{B.}~\bibnamefont{Gehlsen}}, \bibnamefont{and}
  \bibinfo{author}{\bibfnamefont{D.}~\bibnamefont{Helbing}},
  \bibinfo{journal}{Phys. A - Stat. Mech. and Appl.}
  \textbf{\bibinfo{volume}{363}}, \bibinfo{pages}{89} (\bibinfo{year}{2006}).

\bibitem[{\citenamefont{Eppstein and Wang}(2004)}]{JGraphAlgAppl_EW04}
\bibinfo{author}{\bibfnamefont{D.}~\bibnamefont{Eppstein}} \bibnamefont{and}
  \bibinfo{author}{\bibfnamefont{J.}~\bibnamefont{Wang}}, \bibinfo{journal}{J.
  Graph Alg. Appl.} \textbf{\bibinfo{volume}{8}}, \bibinfo{pages}{39}
  (\bibinfo{year}{2004}).

\bibitem[{\citenamefont{Goh et~al.}(2001)\citenamefont{Goh, Kahng, and
  Kim}}]{PhysRevLett_GKK01}
\bibinfo{author}{\bibfnamefont{K.~I.} \bibnamefont{Goh}},
  \bibinfo{author}{\bibfnamefont{B.}~\bibnamefont{Kahng}}, \bibnamefont{and}
  \bibinfo{author}{\bibfnamefont{D.}~\bibnamefont{Kim}},
  \bibinfo{journal}{Phys. Rev. Lett.} \textbf{\bibinfo{volume}{87}},
  \bibinfo{pages}{278701} (\bibinfo{year}{2001}).

\bibitem[{\citenamefont{Newman}(2001)}]{PhysRevE_N01}
\bibinfo{author}{\bibfnamefont{M.~E.~J.} \bibnamefont{Newman}},
  \bibinfo{journal}{Phys. Rev. E} \textbf{\bibinfo{volume}{64}},
  \bibinfo{pages}{016132} (\bibinfo{year}{2001}).

\bibitem[{\citenamefont{Everett and Borgatti}(1999)}]{EverettBorgatti99}
\bibinfo{author}{\bibfnamefont{M.}~\bibnamefont{Everett}} \bibnamefont{and}
  \bibinfo{author}{\bibfnamefont{S.}~\bibnamefont{Borgatti}},
  \bibinfo{journal}{J. of Math. Soc.} \textbf{\bibinfo{volume}{23}},
  \bibinfo{pages}{181} (\bibinfo{year}{1999}).

\bibitem[{\citenamefont{Puzis et~al.}(2007)\citenamefont{Puzis, Elovici, and
  Dolev}}]{Puzis2007}
\bibinfo{author}{\bibfnamefont{R.}~\bibnamefont{Puzis}},
  \bibinfo{author}{\bibfnamefont{Y.}~\bibnamefont{Elovici}}, \bibnamefont{and}
  \bibinfo{author}{\bibfnamefont{S.}~\bibnamefont{Dolev}},
  \bibinfo{journal}{Phys. Rev. E} \textbf{\bibinfo{volume}{76}},
  \bibinfo{pages}{056709} (\bibinfo{year}{2007}).

\bibitem[{\citenamefont{Everett and Borgatti}(2005)}]{SocNet_EB05}
\bibinfo{author}{\bibfnamefont{M.}~\bibnamefont{Everett}} \bibnamefont{and}
  \bibinfo{author}{\bibfnamefont{S.}~\bibnamefont{Borgatti}},
  \bibinfo{journal}{Soc. Networks} \textbf{\bibinfo{volume}{27}},
  \bibinfo{pages}{31} (\bibinfo{year}{2005}).

\bibitem[{\citenamefont{Bonacich}(1972)}]{JMathSoc_B72}
\bibinfo{author}{\bibfnamefont{P.}~\bibnamefont{Bonacich}},
  \bibinfo{journal}{J. Math. Soc.} \textbf{\bibinfo{volume}{2}},
  \bibinfo{pages}{113} (\bibinfo{year}{1972}).

\bibitem[{\citenamefont{Bonacich}(2007)}]{SocNet_B07}
\bibinfo{author}{\bibfnamefont{P.}~\bibnamefont{Bonacich}},
  \bibinfo{journal}{Social Networks} \textbf{\bibinfo{volume}{29}},
  \bibinfo{pages}{555} (\bibinfo{year}{2007}).

\bibitem[{\citenamefont{Noh and Rieger}(2004)}]{PhysRevLett_NR04}
\bibinfo{author}{\bibfnamefont{J.~D.} \bibnamefont{Noh}} \bibnamefont{and}
  \bibinfo{author}{\bibfnamefont{H.}~\bibnamefont{Rieger}},
  \bibinfo{journal}{Phys. Rev. Lett.} \textbf{\bibinfo{volume}{92}},
  \bibinfo{pages}{118701} (\bibinfo{year}{2004}).

\bibitem[{\citenamefont{Newman}(2005)}]{NewmanBCrandwalk}
\bibinfo{author}{\bibfnamefont{M.}~\bibnamefont{Newman}},
  \bibinfo{journal}{Soc. Netw.} \textbf{\bibinfo{volume}{27}},
  \bibinfo{pages}{39} (\bibinfo{year}{2005}).

\bibitem[{\citenamefont{Stephenson and Zelen}(1989)}]{SocNet_SZ89}
\bibinfo{author}{\bibfnamefont{K.}~\bibnamefont{Stephenson}} \bibnamefont{and}
  \bibinfo{author}{\bibfnamefont{M.}~\bibnamefont{Zelen}},
  \bibinfo{journal}{Soc. Networks} \textbf{\bibinfo{volume}{11}},
  \bibinfo{pages}{1} (\bibinfo{year}{1989}).

\bibitem[{\citenamefont{Everett and Borgatti}(2010)}]{SocNet_EB10}
\bibinfo{author}{\bibfnamefont{M.~G.} \bibnamefont{Everett}} \bibnamefont{and}
  \bibinfo{author}{\bibfnamefont{S.~P.} \bibnamefont{Borgatti}},
  \bibinfo{journal}{Soc. Networks} \textbf{\bibinfo{volume}{32}},
  \bibinfo{pages}{339} (\bibinfo{year}{2010}).

\bibitem[{\citenamefont{Ercsey-Ravasz and Toroczkai}(2010)}]{PRLcikkunk}
\bibinfo{author}{\bibfnamefont{M.}~\bibnamefont{Ercsey-Ravasz}}
  \bibnamefont{and}
  \bibinfo{author}{\bibfnamefont{Z.}~\bibnamefont{Toroczkai}},
  \bibinfo{journal}{Phys. Rev. Lett.} \textbf{\bibinfo{volume}{105}},
  \bibinfo{pages}{038701} (\bibinfo{year}{2010}).

\bibitem[{\citenamefont{Arenas et~al.}(2001)\citenamefont{Arenas, Diaz-Guilera,
  and Guimera}}]{PhysRevLett_ADG01}
\bibinfo{author}{\bibfnamefont{A.}~\bibnamefont{Arenas}},
  \bibinfo{author}{\bibfnamefont{A.}~\bibnamefont{Diaz-Guilera}},
  \bibnamefont{and} \bibinfo{author}{\bibfnamefont{R.}~\bibnamefont{Guimera}},
  \bibinfo{journal}{Phys. Rev. Lett.} \textbf{\bibinfo{volume}{86}},
  \bibinfo{pages}{3196} (\bibinfo{year}{2001}).

\bibitem[{\citenamefont{Guimera et~al.}(2002)\citenamefont{Guimera,
  Diaz-Guilera, Vega-Redondo, Cabrales, and Arenas}}]{PhysRevLett_GDVCA02}
\bibinfo{author}{\bibfnamefont{R.}~\bibnamefont{Guimera}},
  \bibinfo{author}{\bibfnamefont{A.}~\bibnamefont{Diaz-Guilera}},
  \bibinfo{author}{\bibfnamefont{F.}~\bibnamefont{Vega-Redondo}},
  \bibinfo{author}{\bibfnamefont{A.}~\bibnamefont{Cabrales}}, \bibnamefont{and}
  \bibinfo{author}{\bibfnamefont{A.}~\bibnamefont{Arenas}},
  \bibinfo{journal}{Phys. Rev. Lett.} \textbf{\bibinfo{volume}{89}},
  \bibinfo{pages}{248701} (\bibinfo{year}{2002}).

\bibitem[{\citenamefont{Yan et~al.}(2006)\citenamefont{Yan, Zhou, Hu, Fu, and
  Wang}}]{PhysRevE_YZHFW06}
\bibinfo{author}{\bibfnamefont{G.}~\bibnamefont{Yan}},
  \bibinfo{author}{\bibfnamefont{T.}~\bibnamefont{Zhou}},
  \bibinfo{author}{\bibfnamefont{B.}~\bibnamefont{Hu}},
  \bibinfo{author}{\bibfnamefont{Z.-Q.} \bibnamefont{Fu}}, \bibnamefont{and}
  \bibinfo{author}{\bibfnamefont{B.-H.} \bibnamefont{Wang}},
  \bibinfo{journal}{Phys. Rev. E} \textbf{\bibinfo{volume}{73}},
  \bibinfo{pages}{046108} (\bibinfo{year}{2006}).

\bibitem[{\citenamefont{Danila et~al.}(2006{\natexlab{a}})\citenamefont{Danila,
  Yu, Earl, Marsh, Toroczkai, and Bassler}}]{Danila}
\bibinfo{author}{\bibfnamefont{B.}~\bibnamefont{Danila}},
  \bibinfo{author}{\bibfnamefont{Y.}~\bibnamefont{Yu}},
  \bibinfo{author}{\bibfnamefont{S.}~\bibnamefont{Earl}},
  \bibinfo{author}{\bibfnamefont{J.~A.} \bibnamefont{Marsh}},
  \bibinfo{author}{\bibfnamefont{Z.}~\bibnamefont{Toroczkai}},
  \bibnamefont{and} \bibinfo{author}{\bibfnamefont{K.~E.}
  \bibnamefont{Bassler}}, \bibinfo{journal}{Phys. Rev. E}
  \textbf{\bibinfo{volume}{74}}, \bibinfo{pages}{046114}
  (\bibinfo{year}{2006}{\natexlab{a}}).

\bibitem[{\citenamefont{Danila et~al.}(2006{\natexlab{b}})\citenamefont{Danila,
  Yu, Marsh, and Bassler}}]{DanilaPRE2006}
\bibinfo{author}{\bibfnamefont{B.}~\bibnamefont{Danila}},
  \bibinfo{author}{\bibfnamefont{Y.}~\bibnamefont{Yu}},
  \bibinfo{author}{\bibfnamefont{J.~A.} \bibnamefont{Marsh}}, \bibnamefont{and}
  \bibinfo{author}{\bibfnamefont{K.~E.} \bibnamefont{Bassler}},
  \bibinfo{journal}{Phys. Rev. E} \textbf{\bibinfo{volume}{74}},
  \bibinfo{pages}{046106} (\bibinfo{year}{2006}{\natexlab{b}}).

\bibitem[{\citenamefont{Danila et~al.}(2007)\citenamefont{Danila, Yu, Marsh,
  and Bassler}}]{DanilaCHAOS2007}
\bibinfo{author}{\bibfnamefont{B.}~\bibnamefont{Danila}},
  \bibinfo{author}{\bibfnamefont{Y.}~\bibnamefont{Yu}},
  \bibinfo{author}{\bibfnamefont{J.~A.} \bibnamefont{Marsh}}, \bibnamefont{and}
  \bibinfo{author}{\bibfnamefont{K.~E.} \bibnamefont{Bassler}},
  \bibinfo{journal}{Chaos} \textbf{\bibinfo{volume}{17}},
  \bibinfo{pages}{026102} (\bibinfo{year}{2007}).

\bibitem[{\citenamefont{Leighton and Rao}(1999)}]{JACM_LR99}
\bibinfo{author}{\bibfnamefont{F.~T.} \bibnamefont{Leighton}} \bibnamefont{and}
  \bibinfo{author}{\bibfnamefont{S.}~\bibnamefont{Rao}}, \bibinfo{journal}{J.
  ACM} \textbf{\bibinfo{volume}{46}}, \bibinfo{pages}{787}
  (\bibinfo{year}{1999}).

\bibitem[{\citenamefont{Vazirani}(2003)}]{Book_Vazirani_03}
\bibinfo{author}{\bibfnamefont{V.~V.} \bibnamefont{Vazirani}},
  \emph{\bibinfo{title}{Approximation Algorithms}}
  (\bibinfo{publisher}{Springer}, \bibinfo{year}{2003}), \bibinfo{edition}{2nd}
  ed.

\bibitem[{\citenamefont{Gkantsidis et~al.}(2003)\citenamefont{Gkantsidis,
  Mihail, and Saberi}}]{SIGMETRICS_GMS03}
\bibinfo{author}{\bibfnamefont{C.}~\bibnamefont{Gkantsidis}},
  \bibinfo{author}{\bibfnamefont{M.}~\bibnamefont{Mihail}}, \bibnamefont{and}
  \bibinfo{author}{\bibfnamefont{A.}~\bibnamefont{Saberi}}, in
  \emph{\bibinfo{booktitle}{Proceedings of the 2003 ACM SIGMETRICS
  international conference on Measurement and modeling of computer systems}}
  (\bibinfo{publisher}{ACM}, \bibinfo{year}{2003}).

\bibitem[{\citenamefont{Akella et~al.}(2003)\citenamefont{Akella, Chawla,
  Kannan, and Sheshan}}]{PODC_ACKS03}
\bibinfo{author}{\bibfnamefont{A.}~\bibnamefont{Akella}},
  \bibinfo{author}{\bibfnamefont{S.}~\bibnamefont{Chawla}},
  \bibinfo{author}{\bibfnamefont{A.}~\bibnamefont{Kannan}}, \bibnamefont{and}
  \bibinfo{author}{\bibfnamefont{S.}~\bibnamefont{Sheshan}}, in
  \emph{\bibinfo{booktitle}{Proceedings of the ACM Principles of Distributed
  Computing (PODC) Conference}} (\bibinfo{address}{Boston, MA},
  \bibinfo{year}{2003}).

\bibitem[{\citenamefont{Dall'Asta
  et~al.}(2006{\natexlab{a}})\citenamefont{Dall'Asta, Alvarez-Hamelin, Barrat,
  V\'azquez, and Vespignani}}]{vespignani-traceroute}
\bibinfo{author}{\bibfnamefont{L.}~\bibnamefont{Dall'Asta}},
  \bibinfo{author}{\bibfnamefont{I.}~\bibnamefont{Alvarez-Hamelin}},
  \bibinfo{author}{\bibfnamefont{A.}~\bibnamefont{Barrat}},
  \bibinfo{author}{\bibfnamefont{A.}~\bibnamefont{V\'azquez}},
  \bibnamefont{and}
  \bibinfo{author}{\bibfnamefont{A.}~\bibnamefont{Vespignani}},
  \bibinfo{journal}{Theor. Comp. Sci.} \textbf{\bibinfo{volume}{355}},
  \bibinfo{pages}{6} (\bibinfo{year}{2006}{\natexlab{a}}).

\bibitem[{\citenamefont{Dall'Asta et~al.}(2005)\citenamefont{Dall'Asta,
  Alvarez-Hamelin, Barrat, V\'azquez, and Vespignani}}]{PRE-int-explor}
\bibinfo{author}{\bibfnamefont{L.}~\bibnamefont{Dall'Asta}},
  \bibinfo{author}{\bibfnamefont{I.}~\bibnamefont{Alvarez-Hamelin}},
  \bibinfo{author}{\bibfnamefont{A.}~\bibnamefont{Barrat}},
  \bibinfo{author}{\bibfnamefont{A.}~\bibnamefont{V\'azquez}},
  \bibnamefont{and}
  \bibinfo{author}{\bibfnamefont{A.}~\bibnamefont{Vespignani}},
  \bibinfo{journal}{Phys. Rev. E} \textbf{\bibinfo{volume}{71}},
  \bibinfo{pages}{036135} (\bibinfo{year}{2005}).

\bibitem[{\citenamefont{Holme et~al.}(2002)\citenamefont{Holme, Kim, Yoon, and
  Han}}]{holme}
\bibinfo{author}{\bibfnamefont{P.}~\bibnamefont{Holme}},
  \bibinfo{author}{\bibfnamefont{B.~J.} \bibnamefont{Kim}},
  \bibinfo{author}{\bibfnamefont{C.~N.} \bibnamefont{Yoon}}, \bibnamefont{and}
  \bibinfo{author}{\bibfnamefont{S.~K.} \bibnamefont{Han}},
  \bibinfo{journal}{Phys. Rev. E} \textbf{\bibinfo{volume}{65}},
  \bibinfo{pages}{056109} (\bibinfo{year}{2002}).

\bibitem[{\citenamefont{Motter and Lai}(2002)}]{PhysRevE_ML02}
\bibinfo{author}{\bibfnamefont{A.~E.} \bibnamefont{Motter}} \bibnamefont{and}
  \bibinfo{author}{\bibfnamefont{Y.~C.} \bibnamefont{Lai}},
  \bibinfo{journal}{Physical Review E} \textbf{\bibinfo{volume}{66}},
  \bibinfo{pages}{065102} (\bibinfo{year}{2002}).

\bibitem[{\citenamefont{Motter}(2004)}]{motter}
\bibinfo{author}{\bibfnamefont{A.~E.} \bibnamefont{Motter}},
  \bibinfo{journal}{Phys. Rev. Lett.} \textbf{\bibinfo{volume}{93}},
  \bibinfo{pages}{098701} (\bibinfo{year}{2004}).

\bibitem[{\citenamefont{Vespignani}(2009)}]{vesp09}
\bibinfo{author}{\bibfnamefont{A.}~\bibnamefont{Vespignani}},
  \bibinfo{journal}{Science} \textbf{\bibinfo{volume}{325}},
  \bibinfo{pages}{425} (\bibinfo{year}{2009}).

\bibitem[{\citenamefont{Dall'Asta
  et~al.}(2006{\natexlab{b}})\citenamefont{Dall'Asta, Barrat, Barthelemy, and
  Vespignani}}]{vespignani}
\bibinfo{author}{\bibfnamefont{L.}~\bibnamefont{Dall'Asta}},
  \bibinfo{author}{\bibfnamefont{A.}~\bibnamefont{Barrat}},
  \bibinfo{author}{\bibfnamefont{M.}~\bibnamefont{Barthelemy}},
  \bibnamefont{and}
  \bibinfo{author}{\bibfnamefont{A.}~\bibnamefont{Vespignani}},
  \bibinfo{journal}{J. Stat. Mech. - Theory and Experiment}
  (\bibinfo{year}{2006}{\natexlab{b}}).

\bibitem[{\citenamefont{Freeman et~al.}(1991)\citenamefont{Freeman, Borgatti,
  and White}}]{SocNet_FBW91}
\bibinfo{author}{\bibfnamefont{L.~C.} \bibnamefont{Freeman}},
  \bibinfo{author}{\bibfnamefont{S.~P.} \bibnamefont{Borgatti}},
  \bibnamefont{and} \bibinfo{author}{\bibfnamefont{D.~R.} \bibnamefont{White}},
  \bibinfo{journal}{Soc. Networks} \textbf{\bibinfo{volume}{13}},
  \bibinfo{pages}{141} (\bibinfo{year}{1991}).

\bibitem[{\citenamefont{Brandes}(2001)}]{JMathSoc_B01}
\bibinfo{author}{\bibfnamefont{U.}~\bibnamefont{Brandes}}, \bibinfo{journal}{J.
  Math. Sociology} \textbf{\bibinfo{volume}{25}}, \bibinfo{pages}{163}
  (\bibinfo{year}{2001}).

\bibitem[{\citenamefont{Barrat et~al.}(2004)\citenamefont{Barrat, Barthelemy,
  Pastor-Satorras, and Vespignani}}]{PNAS_BBPV04}
\bibinfo{author}{\bibfnamefont{A.}~\bibnamefont{Barrat}},
  \bibinfo{author}{\bibfnamefont{M.}~\bibnamefont{Barthelemy}},
  \bibinfo{author}{\bibfnamefont{R.}~\bibnamefont{Pastor-Satorras}},
  \bibnamefont{and}
  \bibinfo{author}{\bibfnamefont{A.}~\bibnamefont{Vespignani}},
  \bibinfo{journal}{Proc.Natl.Acad.Sci. USA} \textbf{\bibinfo{volume}{101}},
  \bibinfo{pages}{3747} (\bibinfo{year}{2004}).

\bibitem[{\citenamefont{Wang et~al.}(2008)\citenamefont{Wang, Hernandez, and
  Mieghem}}]{PhysRevE_WHV08}
\bibinfo{author}{\bibfnamefont{H.}~\bibnamefont{Wang}},
  \bibinfo{author}{\bibfnamefont{J.~M.} \bibnamefont{Hernandez}},
  \bibnamefont{and} \bibinfo{author}{\bibfnamefont{P.~V.}
  \bibnamefont{Mieghem}}, \bibinfo{journal}{Phys. Rev. E}
  \textbf{\bibinfo{volume}{77}}, \bibinfo{pages}{046105}
  (\bibinfo{year}{2008}).

\bibitem[{\citenamefont{Opsahl et~al.}(2010)\citenamefont{Opsahl, Agneessens,
  and Skvoretz}}]{SocNet_OAS10}
\bibinfo{author}{\bibfnamefont{T.}~\bibnamefont{Opsahl}},
  \bibinfo{author}{\bibfnamefont{F.}~\bibnamefont{Agneessens}},
  \bibnamefont{and} \bibinfo{author}{\bibfnamefont{J.}~\bibnamefont{Skvoretz}},
  \bibinfo{journal}{Soc. Networks} \textbf{\bibinfo{volume}{32}},
  \bibinfo{pages}{245 } (\bibinfo{year}{2010}).

\bibitem[{\citenamefont{Granovetter}(1973)}]{AmJSoc_G73}
\bibinfo{author}{\bibfnamefont{M.}~\bibnamefont{Granovetter}},
  \bibinfo{journal}{American Journal of Sociology}
  \textbf{\bibinfo{volume}{78}}, \bibinfo{pages}{1360} (\bibinfo{year}{1973}).

\bibitem[{\citenamefont{Colizza et~al.}(2007)\citenamefont{Colizza,
  Pastor-Satorras, and Vespignani}}]{NatPhys_CPV07}
\bibinfo{author}{\bibfnamefont{V.}~\bibnamefont{Colizza}},
  \bibinfo{author}{\bibfnamefont{R.}~\bibnamefont{Pastor-Satorras}},
  \bibnamefont{and}
  \bibinfo{author}{\bibfnamefont{A.}~\bibnamefont{Vespignani}},
  \bibinfo{journal}{Nature Phys.} \textbf{\bibinfo{volume}{3}},
  \bibinfo{pages}{276} (\bibinfo{year}{2007}).

\bibitem[{\citenamefont{Szabo et~al.}(2002)\citenamefont{Szabo, Alava, and
  Kertesz}}]{PhysRevE_SzAK02}
\bibinfo{author}{\bibfnamefont{G.}~\bibnamefont{Szabo}},
  \bibinfo{author}{\bibfnamefont{M.}~\bibnamefont{Alava}}, \bibnamefont{and}
  \bibinfo{author}{\bibfnamefont{J.}~\bibnamefont{Kertesz}},
  \bibinfo{journal}{Phys. Rev. E} \textbf{\bibinfo{volume}{66}},
  \bibinfo{pages}{026101} (\bibinfo{year}{2002}).

\bibitem[{\citenamefont{Bollob\'as and Riordan}(2004)}]{PhysRevE_BR04}
\bibinfo{author}{\bibfnamefont{B.}~\bibnamefont{Bollob\'as}} \bibnamefont{and}
  \bibinfo{author}{\bibfnamefont{O.}~\bibnamefont{Riordan}},
  \bibinfo{journal}{Phys. Rev. E} \textbf{\bibinfo{volume}{69}},
  \bibinfo{pages}{036114} (\bibinfo{year}{2004}).

\bibitem[{\citenamefont{Fekete et~al.}(2006)\citenamefont{Fekete, Vattay, and
  Kocarev}}]{PhysRevE_FVK06}
\bibinfo{author}{\bibfnamefont{A.}~\bibnamefont{Fekete}},
  \bibinfo{author}{\bibfnamefont{G.}~\bibnamefont{Vattay}}, \bibnamefont{and}
  \bibinfo{author}{\bibfnamefont{L.}~\bibnamefont{Kocarev}},
  \bibinfo{journal}{Phys. Rev. E} \textbf{\bibinfo{volume}{73}},
  \bibinfo{pages}{046102} (\bibinfo{year}{2006}).

\bibitem[{\citenamefont{Kitsak et~al.}(2007)\citenamefont{Kitsak, Havlin, Paul,
  Riccaboni, Pammolli, and Stanley}}]{PhysRevE_KHPRPS07}
\bibinfo{author}{\bibfnamefont{M.}~\bibnamefont{Kitsak}},
  \bibinfo{author}{\bibfnamefont{S.}~\bibnamefont{Havlin}},
  \bibinfo{author}{\bibfnamefont{G.}~\bibnamefont{Paul}},
  \bibinfo{author}{\bibfnamefont{M.}~\bibnamefont{Riccaboni}},
  \bibinfo{author}{\bibfnamefont{F.}~\bibnamefont{Pammolli}}, \bibnamefont{and}
  \bibinfo{author}{\bibfnamefont{H.~E.} \bibnamefont{Stanley}},
  \bibinfo{journal}{Phys. Rev. E} \textbf{\bibinfo{volume}{75}},
  \bibinfo{pages}{056115} (\bibinfo{year}{2007}).

\bibitem[{\citenamefont{Johnson}(1977)}]{johnson}
\bibinfo{author}{\bibfnamefont{D.~B.} \bibnamefont{Johnson}},
  \bibinfo{journal}{J. ACM} \textbf{\bibinfo{volume}{24}}, \bibinfo{pages}{1}
  (\bibinfo{year}{1977}), ISSN \bibinfo{issn}{0004-5411}.

\bibitem[{\citenamefont{Floyd}(1962)}]{floyd}
\bibinfo{author}{\bibfnamefont{R.~W.} \bibnamefont{Floyd}},
  \bibinfo{journal}{Commun. ACM} \textbf{\bibinfo{volume}{5}},
  \bibinfo{pages}{345} (\bibinfo{year}{1962}).

\bibitem[{\citenamefont{Warshall}(1962)}]{warshall}
\bibinfo{author}{\bibfnamefont{S.}~\bibnamefont{Warshall}},
  \bibinfo{journal}{J. ACM} \textbf{\bibinfo{volume}{9}}, \bibinfo{pages}{11}
  (\bibinfo{year}{1962}), ISSN \bibinfo{issn}{0004-5411}.

\bibitem[{\citenamefont{Brandes and Pich}(2007)}]{brandes-approx}
\bibinfo{author}{\bibfnamefont{U.}~\bibnamefont{Brandes}} \bibnamefont{and}
  \bibinfo{author}{\bibfnamefont{C.}~\bibnamefont{Pich}}, \bibinfo{journal}{I.
  J. Bif. and Chaos} \textbf{\bibinfo{volume}{17}}, \bibinfo{pages}{2303}
  (\bibinfo{year}{2007}).

\bibitem[{\citenamefont{Geisberger et~al.}(2008)\citenamefont{Geisberger,
  Sanders, and Schultes}}]{geisberger-approx}
\bibinfo{author}{\bibfnamefont{R.}~\bibnamefont{Geisberger}},
  \bibinfo{author}{\bibfnamefont{P.}~\bibnamefont{Sanders}}, \bibnamefont{and}
  \bibinfo{author}{\bibfnamefont{D.}~\bibnamefont{Schultes}}, in
  \emph{\bibinfo{booktitle}{ALENEX}} (\bibinfo{year}{2008}), pp.
  \bibinfo{pages}{90--100}.

\bibitem[{\citenamefont{Gonz\'alez et~al.}(2008)\citenamefont{Gonz\'alez,
  Hidalgo, and Barab\'asi}}]{Barabasi-mobility}
\bibinfo{author}{\bibfnamefont{M.~C.} \bibnamefont{Gonz\'alez}},
  \bibinfo{author}{\bibfnamefont{C.~A.} \bibnamefont{Hidalgo}},
  \bibnamefont{and} \bibinfo{author}{\bibfnamefont{A.-L.}
  \bibnamefont{Barab\'asi}}, \bibinfo{journal}{Nature}
  \textbf{\bibinfo{volume}{453}}, \bibinfo{pages}{779} (\bibinfo{year}{2008}).

\bibitem[{\citenamefont{Onnela et~al.}(2007)\citenamefont{Onnela, Saramaki,
  Hyvonen, Szabo, Lazer, Kaski, Kertesz, and Barabasi}}]{PNAS_OSHSLKKB07}
\bibinfo{author}{\bibfnamefont{J.}~\bibnamefont{Onnela}},
  \bibinfo{author}{\bibfnamefont{J.}~\bibnamefont{Saramaki}},
  \bibinfo{author}{\bibfnamefont{J.}~\bibnamefont{Hyvonen}},
  \bibinfo{author}{\bibfnamefont{G.}~\bibnamefont{Szabo}},
  \bibinfo{author}{\bibfnamefont{D.}~\bibnamefont{Lazer}},
  \bibinfo{author}{\bibfnamefont{K.}~\bibnamefont{Kaski}},
  \bibinfo{author}{\bibfnamefont{J.}~\bibnamefont{Kertesz}}, \bibnamefont{and}
  \bibinfo{author}{\bibfnamefont{A.}~\bibnamefont{Barabasi}},
  \bibinfo{journal}{Proc.Natl.Acad.Sci. USA} \textbf{\bibinfo{volume}{104}},
  \bibinfo{pages}{7332} (\bibinfo{year}{2007}).

\bibitem[{\citenamefont{Newman et~al.}(2001)\citenamefont{Newman, Strogatz, and
  Watts}}]{newman_randomgraphs}
\bibinfo{author}{\bibfnamefont{M.~E.~J.} \bibnamefont{Newman}},
  \bibinfo{author}{\bibfnamefont{S.~H.} \bibnamefont{Strogatz}},
  \bibnamefont{and} \bibinfo{author}{\bibfnamefont{D.~J.} \bibnamefont{Watts}},
  \bibinfo{journal}{Phys. Rev. E} \textbf{\bibinfo{volume}{64}},
  \bibinfo{pages}{026118} (\bibinfo{year}{2001}).

\bibitem[{\citenamefont{Shao et~al.}(2009)\citenamefont{Shao, Buldyrev,
  Braunstein, Havlin, and Stanley}}]{shell_structure}
\bibinfo{author}{\bibfnamefont{J.}~\bibnamefont{Shao}},
  \bibinfo{author}{\bibfnamefont{S.~V.} \bibnamefont{Buldyrev}},
  \bibinfo{author}{\bibfnamefont{L.~A.} \bibnamefont{Braunstein}},
  \bibinfo{author}{\bibfnamefont{S.}~\bibnamefont{Havlin}}, \bibnamefont{and}
  \bibinfo{author}{\bibfnamefont{H.~E.} \bibnamefont{Stanley}},
  \bibinfo{journal}{Phys. Rev. E} \textbf{\bibinfo{volume}{80}},
  \bibinfo{pages}{036105} (\bibinfo{year}{2009}).

\bibitem[{\citenamefont{Penrose}(2003)}]{penrose}
\bibinfo{author}{\bibfnamefont{M.}~\bibnamefont{Penrose}},
  \emph{\bibinfo{title}{Random Geometric Graphs (Oxford Studies in
  Probability)}} (\bibinfo{publisher}{Oxford University Press, USA},
  \bibinfo{year}{2003}), ISBN \bibinfo{isbn}{0198506260}.

\bibitem[{\citenamefont{Dall and Christensen}(2002)}]{RG}
\bibinfo{author}{\bibfnamefont{J.}~\bibnamefont{Dall}} \bibnamefont{and}
  \bibinfo{author}{\bibfnamefont{M.}~\bibnamefont{Christensen}},
  \bibinfo{journal}{Phys. Rev. E} \textbf{\bibinfo{volume}{66}},
  \bibinfo{pages}{016121} (\bibinfo{year}{2002}).

\bibitem[{\citenamefont{Serrano et~al.}(2009)\citenamefont{Serrano, Boguna, and
  Vespignani}}]{PNAS_SBV09}
\bibinfo{author}{\bibfnamefont{M.~A.} \bibnamefont{Serrano}},
  \bibinfo{author}{\bibfnamefont{M.}~\bibnamefont{Boguna}}, \bibnamefont{and}
  \bibinfo{author}{\bibfnamefont{A.}~\bibnamefont{Vespignani}},
  \bibinfo{journal}{PNAS} \textbf{\bibinfo{volume}{106}}, \bibinfo{pages}{6483}
  (\bibinfo{year}{2009}).

\end{thebibliography}

\end{document}